\documentstyle[aps,preprint,eqsecnum]{revtex}
\draft
\tightenlines

\makeatletter
                \def\@preprint{}
                \def\preprint#1#2#3#4{%
                \ifpreprintsty
                \def\@preprint{
                \noindent \hbox{#1}\hfill\hbox{#2}\\
                \hbox{#3}\hfill\hbox{#4} \vskip 8pt}%
                \fi
                }
\makeatother

\begin{document}

\preprint{hep-ph/9704406}{CERN-TH/97-80}{}{NTZ 12/97}

\title{Restricted conformal invariance in QCD and its
predictive power for virtual two-photon processes}
\author{
        D. M\"uller
        }
\address{TH Division, CERN, 1211 Geneva 23, Switzerland
        \footnote{
			On leave from the Institut f\"ur Theoretische Physik,
			Universit\"at Leipzig, 04109 Leipzig, Germany.}
        }

\date{23.12.98}
\maketitle

\begin{abstract}
The conformal algebra provides powerful constraints, which guarantee that
renormalized conformally covariant operators exist in the hypothetical
conformal limit of the theory, where the $\beta$-function vanishes. Thus, in
this limit also the conformally covariant operator product expansion on the
light cone holds true. This operator product expansion has predictive power
for two-photon processes in the generalized Bjorken region. Only the Wilson
coefficients and the anomalous dimensions that are known from deep inelastic
scattering are required for the prediction of all other two-photon processes
in terms of the process-dependent off-diagonal expectation values of
conformal operators. It is checked that the next-to-leading order
calculations for the flavour non-singlet meson transition form factors are
consistent with the corrections to the corresponding Wilson coefficients in
deep inelasitic scattering.
\end{abstract}

\newpage
\narrowtext

\section{Introduction}

In a massless theory, conformal symmetry has the ability to provide powerful
predictions for physical quantities. However, conformal symmetry is broken
on the quantum level due to the renormalization. This breaking is controlled
by the conformal Ward identities (CWI) of dilatation and special conformal
symmetry. It is often possible to redefine the conformal representation,
implying that in the physical sector of the theory the conformal symmetry is
only broken by the renormalization of the running coupling constant, which
provides a symmetry-breaking term proportional to the $\beta$-function. If
there exists a non-trivial (non-perturbative) fixed-point such that the
$\beta$-function vanishes, conformal invariance holds true. This is the
so-called hypothetical conformal limit of the theory.

For instance, the Crewther relation \cite{Cre72} (also derived in
\cite{AdlCalGroJac72}) is based on conformal invariance. The value of the
Bjorken sum rule $K$ and the isovector part $R'$ of the cross-section ratio
$\sigma(e^+ e^- \to \mbox{hadrons})/\sigma(e^+ e^- \to \mu^+ \mu^-)$ are
computed in QCD up to the order $\alpha_s^3$ \cite{LarVer91,GorKatLar91}.
Assuming conformal invariance for the axial vector-vector-vector (AVV)
correlator, Crewther proved that
the relation $3S =K R'$ holds true; $S$ is the anomalous constant, which is,
corresponding to the Adler--Bardeen theorem, given by its one-loop value.
Since the vector current and the flavour non-singlet axial-vector current
are conserved, the CWI for the AVV correlator tells us that conformal
invariance can be broken only by the $\beta$-function \cite{Cre97}. This is
actually the case for the available $\alpha_s^3$ order \cite{BroKat93} (see
also \cite{BroLu95}--\cite{Rat96}).

It has been known for a long time that the conformal symmetry provides
powerful constraints for the Wilson coefficients of the operator product
expansion (OPE) for two local currents. This provides an improved OPE, which
was called conformally covariant OPE and was studied in a number of papers,
starting in the seventies with the work of Ferrara, Gatto and Grillo
\cite{FerGriGat71,FerGriGat71a,FerGriGat72a,BonSarTon72,CraDobTod85}.
Employing the conformal Ward identities for the Green functions of composite
operators, the conformal symmetry breaking in the interacting scalar theory
was studied by Ferrara, Grillo and Parisi \cite{FerGriPar73}. Making a
non-trivial assumption about the form of the conformal Ward identities, the
authors found that for a non-trivial fixed-point of the $\beta$-function the
conformally covariant OPE holds true if the original scale dimensions, given
by the canonical dimensions, are shifted by the anomalous dimensions.

Brodsky, Frishman, Lepage and Sachradja employed the conformally covariant
expansion for a non-local operator, appearing in the definition of the pion
distribution amplitude, to predict the evolution of this amplitude correctly
at leading order (LO) \cite{BroFriLepSac80} (see also \cite{CraDobTod85}).
Note that this requires a generalization of the OPE and an assumption about
the conformal properties of this non-local operator. With this application a
puzzle of the conformal invariance in gauge field theories sets up beyond
the leading order. The next-to-leading order (NLO) prediction for the
eigenfunctions of the pion evolution kernel in the conformal limit
contradicts the NLO calculation of the non-singlet evolution kernel in the
modified minimal subtraction ($\overline{\rm MS}$) scheme
\cite{DitRad84,Sar84,Kat85,MikRad85}. Brodsky, Damgaard, Frishman and Lepage
have shown that this breakdown of conformal symmetry is
renormalization-scheme-dependent and that, in the special case of the
$\varphi^3_{(6)}$-theory in six dimensions, conformal symmetry is preserved
in the Pauli--Villars regularization at the considered order
\cite{BroDamFriLep86}. However, in QCD the conformal symmetry predictions
could not be restored by a renormalization group transformation. Later, the
breaking of conformal covariance for local conformal composite operators,
appearing in the expansion of the considered non-local operator, was
analysed by the CWI and conformal constraints. It was found that the
symmetry breaking in the $\overline{\rm MS}$ scheme (i) destroys the
irreducible representation of the conformal algebra at {\em leading order}
and (ii) that the conformal symmetry-breaking at this order provides the
eigenfunctions of the computed evolution kernels in NLO \cite{Mue91a,Mue94},
also for QCD.

In this paper we do not reanalyse this issue in terms of a modified
conformal OPE, which may be done in a straightforward manner by employing
the CWI. Here we deal with the conformally covariant OPE for two local
currents that have a well-known behaviour under conformal transformations in
the interacting theory. Neglecting the conformal symmetry-breaking terms
that are proportional to the $\beta$-function allows us to give a first
application to QCD processes.

The paper is organized as follows. For the convenience of the reader, in
Section \ref{Conf-Rev} we shortly review the conformal algebra and the
irreducible representations that are needed, as well as the derivation of
the conformally covariant OPE. Using conformal constraints, coming from the
conformal algebra and the CWI, it will be proved in Section \ref{Conf-Ren}
that there exists a renormalization scheme in which the conformal covariance
of composite operators holds true in the conformal limit. This allows the
construction of the conformally covariant OPE in the interacting theory. In
Section \ref{Conf-Pre} we employ this conformally covariant OPE to predict
the scattering amplitude for two-photon processes in the
light-cone-dominated region \cite{MueRobGeyDitHor94} covering the kinematics
of deeply virtual Compton scattering (DVCS), which has recently been
proposed to open a new window for the exploration of the nucleon contents
\cite{Ji96,Ji97,Rad96}. Included are two well-known special cases,
which have been measured: deep inelastic scattering (DIS) and one-meson
production at large momentum transfer in two-photon collision
\cite{Pluto84,TPC90,CELLO91,Sav97}. In Section \ref{Conf-Appl} we show
that the existing NLO corrections for the flavour non-singlet meson
transition form factors, computed in the $\overline{\rm MS}$ scheme,
coincide with the prediction of the conformally covariant OPE. Finally, the
knowledge of the higher-order corrections to the Bjorken sum rule allows us
to point out phenomenological consequences for the pion transition form
factor.

\section{Conformal algebra and the conformally covariant OPE}
\label{Conf-Rev}

\subsection{Field theoretical conformal representations}

The conformal group is the maximal extension of the Poincar\'e group that
leaves the light cone invariant; it is thus of physical interest for
light-cone-dominated as well as for high-energy processes. Beside the
well-known Poincar\'e transformations, the conformal group consists of
dilatations: $x^\alpha \to \rho x^\alpha$ and special conformal
transformations: $x^\alpha \to (x^\alpha+c^\alpha x^2)/(1+2cx+c^2 x^2)$. The
latter is composed of an inversion $x^\alpha\to x^\alpha/x^2$, a translation
$x^\alpha\to x^\alpha + c^\alpha$, and a further inversion. The conformal
factor $1/(1+2cx+c^2 x^2)$ is singular on the cone and so the special
conformal transformations are not well defined as global transformations in
the Minkowski space. Moreover, it is possible to transform non-causal
connected regions into one another, which violates the principle of
causality. To apply the conformal group to the quantum field theory in
Minkowski space, it is sufficient in future studies to restrict ourselves to
infinitesimal special conformal transformations: this avoids both of the
mentioned problems. In the following we consider field theories (of
polynomial form) with space-time dimensions larger than 2, which are
conformally invariant on the classical level.

For space-time dimension $n$, the algebra of the conformal group is
isomorphic to the algebra $SO(n,2)$ and consists, besides of the algebra of
the Poincar\'e generators $M_{\alpha\beta}$ and $P_{\alpha}$:
\begin{eqnarray}
\label{PoinAlg}
\left[M_{\alpha\beta},M_{\gamma\delta}\right]&=&
	i\left(
	-g_{\alpha\gamma} M_{\beta\delta} + g_{\alpha\delta} M_{\beta\gamma} +
	 g_{\beta\gamma} M_{\alpha\delta} - g_{\beta\delta} M_{\alpha\gamma}
	\right),
\\
\left[M_{\alpha\beta},P_{\gamma}\right] &=&
	i\left(
	-g_{\alpha\gamma} P_{\beta} + g_{\beta\gamma} P_{\alpha}
	\right),
\quad
\left[P_{\alpha},P_{\beta}\right]=
	0,
\nonumber
\end{eqnarray}
of the following commutation relations for the dilatation generator $D$
and for the generator $K_{\alpha}$ of special conformal
transformations:
\begin{eqnarray}
\label{ConfAlg}
\left[D,K_{\alpha}\right] &=&
	i K_{\alpha}, \quad
\left[K_{\alpha},P_{\beta}\right] =
	-2i \left(
			g_{\alpha\beta} D + M_{\alpha\beta}
	    \right),\quad
\left[K_{\alpha},K_{\beta}\right]=0,
\\
\left[D,P_{\alpha}\right] &=&
	-i P_{\alpha}, \quad
\left[M_{\alpha\beta},K_{\gamma}\right] =
	i \left(
			g_{\beta\gamma} K_{\alpha} - g_{\alpha\gamma} K_{\beta} 
	    \right),\quad
\left[D,M_{\alpha\beta}\right]=0.
\nonumber
\end{eqnarray}
The field theoretical representations with finite components
have been classified on the basis of the induced representation
theory \cite{MacSal69}.
In the following we deal with irreducible representations of the conformal
algebra, where the action of  the special conformal generator $K_{\lambda}$
on a basis field $\phi(x)$ vanishes at the point $x^\alpha=0$:
\begin{eqnarray}
\label{ConfLittAlg}
[\phi(x),M_{\alpha\beta}]_{|x=0}=\Sigma_{\alpha\beta}\phi(0),
\quad
[\phi(x),D]_{|x=0} =id_\phi\phi(0), \quad
[\phi(x),K_{\alpha}]_{|x=0}=0.
\end{eqnarray}
Here $\phi={\varphi,\psi,A^\mu}$ may be a scalar field $\varphi$, a
fermionic field $\psi$ or a gauge field $A^\mu$ with scale dimension
$d_\phi$. Conformal invariance in the classical theory requires that the
fields are massless and that the scale dimensions are equal to the canonical
dimensions $d_\phi^{\rm can}$ of the fields. The representation of the
Lorentz group is:
\begin{eqnarray}
\label{litalgLor}
\Sigma_{\alpha\beta}\varphi=0,\quad
\Sigma_{\alpha\beta}\psi=
	\frac{i}{4}\left[\gamma_\alpha,\gamma_\beta\right]\psi,
\quad
\Sigma_{\alpha\beta}A_\mu=
	i\left(g_{\mu\alpha} A_\beta - g_{\mu\beta} A_\alpha \right).
\end{eqnarray}
The theory of induced representation now provides the action of the generators
at an arbitrary space-time point:
\begin{eqnarray}
\label{PoincareGener}
[\phi(x),P_{\alpha}]&=&
	i\partial_{\alpha}\phi(x), \quad
[\phi(x),M_{\alpha\beta}]=
	\left(
		i \left[x_{\alpha} \partial_{\beta}- x_{\beta} \partial_{\alpha}
			\right] 
		+\Sigma_{\alpha\beta}
	\right)\phi(x),
\\
{}[\phi(x),D] &=&
	i(d_\phi+x\partial)\phi(x), \quad
[\phi(x),K_{\alpha}]=
	i \left(
		2 x_{\alpha} (d_\phi+x\partial) -
		2i \Sigma_{\alpha\beta} x^{\beta}-x^2 \partial_{\alpha}
	  \right)\phi(x).
\nonumber
\end{eqnarray}

\subsection{Conformally covariant operators}

Parton distribution functions and hadron distribution amplitudes are defined
as expectation values of composite light-ray operators. Their moments are
expressed in terms of local operators. In the following, the local conformal
two-particle operators are considered. For these operators the conformal
symmetry at tree level yields that the operators do not mix under
renormalization at LO and so they are essential in solving the
evolution equations in the non-forward case. For the leading twist operators
it is sufficient to consider the collinear conformal algebra, which is
isomorphic to $SU(1,1)\equiv SO(2,1)$. The four generators are obtained by
projection onto the light cone:
\begin{eqnarray}
\label{Gen-colconfAlg}
P_+=\tilde{n}_\alpha P^\alpha, \quad
K_-=\tilde{n}^\star_\alpha K^\alpha, \quad
D, \quad
M_{-+}=\tilde{n}^\star_\alpha M^{\alpha\beta} \tilde{n}_\beta,
\end{eqnarray}
where $\tilde{n},\tilde{n}^\star$ are light-cone vectors with the
normalization $\tilde{n}\tilde{n}^\star=1$. They generate the projective
transformation onto the line and satisfy the commutation relation
[coming from the algebra (\ref{ConfAlg})]:
\begin{eqnarray}
\label{ConfAlgCol}
{}[M_{-+},K_-]&=&[D,K_-]=i K_-, \qquad [K_-,P_+] = -2i(D+M_{-+}),
\\
{}[M_{-+},P_+]&=&[D,P_+]=-iP_+, \qquad [M_{-+},D]=0.
\nonumber
\end{eqnarray}
Acting with the step-up operator $P_+$ on $\phi(0)$ generates an
infinite-dimensional representation, the so-called conformal tower
$\{\phi(0),P_+\phi(0), P^2_+\phi(0), \dots\}$. The operator $K_-$ acts as a
step-down one, and annihilates the lowest member $\phi(0)$, i.e.
$K_- \phi(0)=0$. As for $D$ and $M_{-+}$, they are diagonal operators, which
give the scale dimension and the spin of the members, respectively. The
spectrum of the Casimir operator
\begin{eqnarray}
\label{ConfAlgCol-Casimir}
C=\frac{1}{2} P_+K_- -\frac{1}{4}(D+M_{-+})^2 -\frac{i}{2} (D+M_{-+})
\end{eqnarray}
is $j(j+1)$, where $j$ is the conformal spin.

Conformal composite operators can be constructed by different methods
\cite{Mak81,Ohr82,BalBra89}, for instance by decomposition of the direct
product of two towers $\{\phi_i(0),P_+\phi_i(0),\dots\}$
into irreducible representations. The Clebsch-Gordon coefficients are
given by the coefficients of the Jacobi polynomials
$P^{\left(\nu_1-\frac{1}{2},\nu_2-\frac{1}{2}\right)}_k$,
\begin{eqnarray}
\label{confOpe}
O_{kl}(\nu_1,\nu_2)=
	(i\tilde{n}\partial_+)^l \phi_1(0) \Gamma
		P_k^{\left(\nu_1-\frac{1}{2},\nu_2-\frac{1}{2}\right)}\left(
				\frac{\tilde{n}\partial_-}{\tilde{n}\partial_+}
				\right)
		\phi_2(0),\quad l\ge k, \mbox{\ } \nu_i=d_{\phi_i}+s_{\phi_i}-1/2,
\end{eqnarray}
where $\partial_\pm^\nu = 
\stackrel{\rightarrow}{\partial^\nu} \pm \stackrel{\leftarrow}{\partial^\nu}$
and $\Gamma$ contains the spin and, eventually, further global group
theoretical structures. To ensure gauge invariance in a given gauge field
theory, the partial derivatives have to be replaced by the covariant ones,
which does not spoil the conformal properties of the operator.  The index
$\nu_i$ is determined by both the scale dimension and the spin of the
corresponding field $\phi_i$. The action of the collinear conformal
generators are given by:
\begin{eqnarray}
\label{ActOnConfOpe}
\left[O_{kl}(\nu_1,\nu_2),P_+\right] &=& O_{kl+1}(\nu_1,\nu_2)
\nonumber\\
\left[O_{kl}(\nu_1,\nu_2),K_-\right] &=& 2(k-l)(k+l+\nu_1+\nu_2)
					O_{kl-1}(\nu_1,\nu_2),
\\
\left[O_{kl}(\nu_1,\nu_2),D\right] &=& i(d_1+d_2+l) O_{kl}(\nu_1,\nu_2),
\nonumber\\
\left[O_{kl}(\nu_1,\nu_2),M_{-+}\right] &=& i(s_1+s_2+l)
O_{kl}(\nu_1,\nu_2),
\nonumber
\end{eqnarray}
where $(d_1+d_2+l)$ and $(s_1+s_2+l)$ are the dimension and the spin,
respectively,  of the operator $O_{kl}(\nu_1,\nu_2)$.
The lowest member in each tower is $O_{kk}(\nu_1,\nu_2)$.

In the following we often consider conformal operators in the
$\varphi^3_{(6)}$ theory, where the field has the canonical dimension
$d_\varphi=2$, and in QCD for the non-singlet channel, where the quark
fields have $d_\psi=3/2$ and $s=1/2$, so that in both cases
$\nu_1=\nu_2=3/2$. Since $\nu_1=\nu_2$, the Jacobi polynomials can be
expressed by the Gegenbauer polynomials $C_k^\nu$:
\begin{eqnarray}
\label{JacPol-by-GegPol}
P_k^{\left(\nu-\frac{1}{2},\nu-\frac{1}{2}\right)}(x) =
	\frac{(\nu+1/2)_k}{(2\nu)_k}
	C_k^\nu(x),
\end{eqnarray}
where $(a)_n=\Gamma(a+n)/\Gamma(a)$ is the Pochhammer symbol.

\subsection{Conformally covariant OPE}

Here we refer to the work of Ferrara,  Gatto  and Grillo
\cite{FerGriGat71,FerGriGat71a,FerGriGat72a}. The construction of the
conformally covariant OPE is based on the behaviour of the conformal
operators under infinitesimal conformal transformations, which is
characterized by their scaling dimension and their conformal spin. To
simplify the notation, only the scalar case will be considered. In
Subsection \ref{ConRenSch-inttheo} the derivation reviewed here will be
applied to the product of two electromagnetic currents. We restrict
ourselves to leading twist and assume that the following symmetrized and
traceless conformal operators with scale dimension $l_k+l-k=d_1+d_2+l$ form
a complete basis:
\begin{eqnarray}
\label{OPE-def-Ope}
O_{\alpha_1\dots\alpha_l,k}(x)=
\hspace{1mm}
{{\rm S}_{\displaystyle\mbox{\ }_{\mbox{\ }_{\hspace{-6.2mm}\{\alpha\}}}}}\;
	i^{(l-k)} \partial_{\alpha_{k+1}}\dots\partial_{\alpha_l}
	O_{\alpha_1\dots\alpha_k}(x) -\mbox{traces}.
\end{eqnarray}
Here
${{\rm S}_{\displaystyle\mbox{\ }_{\mbox{\ }_{\hspace{-6.2mm}\{\alpha\}}}}}$
denotes symmetrization with respect to $\alpha_1,\dots,\alpha_l$. As in the
case of the collinear conformal representation, which is obtained by
contraction with the light-cone vector $\tilde{n}$, the operator
$O_{\alpha_1\dots\alpha_k}(0)$ is the lowest member of the corresponding
conformal tower. For dimensional reasons, the product of the currents $A$
and $B$ are expanded on the light cone as:
\begin{eqnarray}
\label{OPE-def}
A(x) B(0) &=&
	\sum_{k=0}^\infty
	\left(\frac{1}{x^2}\right)^{\frac{l_A+l_B-l_k+k}{2}}\sum_{l=k}^\infty
	\tilde{C}_{kl} (-i)^{(l+1)} x^{\alpha_1}\dots x^{\alpha_l}
	O_{\alpha_1\dots\alpha_l,k}(0),
\end{eqnarray}
where $\tilde{C}_{kl}$ are the Wilson coefficients. Furthermore, $l_A$ and
$l_B$ denote the scale dimensions of the currents $A$ and $B$, respectively.

One way to construct the conformally covariant OPE is to act with
$K_\lambda$ on both sides of the OPE (\ref{OPE-def}) (see
\cite{FerGriGat71a}) and compare the two results for the leading twist
contributions. Taking into account the action of $K_\lambda$ on the currents
\begin{eqnarray}
\label{cOPE-act-K-cur}
[A(x)B(0),K_\lambda]= 	i \left(
		2 x_{\lambda} (l_A+x\partial) -x^2 \partial_{\lambda}
	  \right) A(x)B(0)
\end{eqnarray}
and on the composite operators
(\ref{OPE-def-Ope}),
\begin{eqnarray}
\label{cOPE-act-K}
\hspace{3mm}
{{\rm S
}_{\displaystyle\mbox{\ }_{\mbox{\ }_{\hspace{-7.2mm}\{\alpha,\lambda\}}}}}
	[O_{\alpha_1\dots\alpha_l,k}(0),K_\lambda]&=&
	2(k-l)(l_k+l-1) \hspace{3mm}
{{\rm S
}_{\displaystyle\mbox{\ }_{\mbox{\ }_{\hspace{-7.2mm}\{\alpha,\lambda\}}}}}
	O_{\alpha_1\dots\alpha_{l-1},k}(0)
	g_{\lambda\alpha_{l}}
\end{eqnarray}
results, after comparison of the obtained expressions, into a recurrence
relation for the Wilson coefficients:
\begin{eqnarray}
\label{cOPE-recrel}
\tilde{C}_{kl+1}=
	\frac{(l_A-l_B+l_k-k)/2+l}{(l-k+1)(l_k+l)}  \tilde{C}_{kl}
\; \Rightarrow \;
\tilde{C}_{kl} =
	\frac{([l_A-l_B+l_k+k]/2)_{l-k}}{(l-k)!
	(l_k+k)_{l-k}} \tilde{C}_{kk}.
\end{eqnarray}
Inserting this solution in the OPE (\ref{OPE-def}) allows 
the summation with respect to $l$ and provides the conformally covariant
OPE that is written here in the following representation:
\begin{eqnarray}
\label{cOPE-def}
A(x) B(0) &=&
	\sum_{k=0}^\infty
	\tilde{C}_k \left(\frac{1}{x^2}\right)^{\frac{l_A+l_B-l_k+k}{2}}
	(-i)^{(k+1)} x^{\alpha_1}\dots x^{\alpha_k}\times
\\
	&&\int_0^1du\, u^{(l_A-l_B+k+l_k)/2-1}
		(1-u)^{(l_B-l_A+k+l_k)/2-1}
	O_{\alpha_1\dots\alpha_k}(ux),
\nonumber
\end{eqnarray}
where $\tilde{C}_k=\tilde{C}_{kk} $.

Let us recall the assumptions used to derive this conformally covariant
OPE (\ref{cOPE-def}). Besides the completeness  of the operator basis, it
was essential that $K_\lambda$ annihilates the currents $A,B$ and
the conformal operators at the point $x_\alpha=0$:
\begin{eqnarray}
\label{cOPE-condCur}
	[A(x),K_\lambda]_{|x=0}=[B(x),K_\lambda]_{|x=0}=0,
\\
\label{cOPE-condOpe}
{{\rm S
}_{\displaystyle\mbox{\ }_{\mbox{\ }_{\hspace{-7.2mm}\{\alpha,\lambda\}}}}}
	[O_{\alpha_1\dots\alpha_k}(x),K_\lambda]_{|x=0}=0.
\end{eqnarray}
The transformation laws under the infinitesimal conformal transformation
(\ref{cOPE-act-K}), analogous to those in (\ref{ActOnConfOpe}), result in a
special arrangement of the operators. This causes the predictive power of
the conformally covariant OPE, namely that the corresponding Wilson
coefficients are already fixed and only the coefficients $\tilde{C}_k$ are
unknown and have to be computed explicitly, which can be done by forming
forward matrix elements. In this case the $u$-dependence of the operators
can be dropped and the conformal OPE is reduced to the common OPE for
the forward case that is familiar from deep inelastic scattering (DIS).

\section{Conformally covariant renormalization scheme}
\label{Conf-Ren}

Generally, the conformal invariance of classical field theories is broken at
the quantum level owing to the renormalization of the fields and the
coupling constant\footnote{In gauge field theories, the conformal invariance
is also broken by the renormalization of the gauge-fixing parameter as well
as explicitly by the gauge-fixing and ghost terms in the action. In the
Abelian theory, this breaking can be formally written as a
Becchi--Rouet--Stora--Tyutin (BRST)
transformation, so that it does not appear in the physical sector of the
theory. In the following we assume that this breaking is also absent in the
physical sector of QCD.}. However, the symmetry breaking by the
renormalization of the fields can be absorbed into the redefinition of the
conformal representation, i.e. the scale dimension given originally by the
canonical dimension is shifted by the anomalous dimension of the
corresponding field. The renormalization of the coupling constant cannot be
implemented in the original irreducible representation. However, in a scalar
theory, Zaikov explored the possibility to extend the conformal
representation of Green functions to a non-decomposable irreducible
representation that includes the $\beta$-function \cite{Zai88}. In the
following, it is simply assumed that there exists a non-trivial fixed-point
such that the $\beta$-function vanishes; formally, we speak from the
conformal limit and set $\beta$ to zero.

The conformal properties of composite operators will also be spoiled by the
renormalization. To study this symmetry breaking we employ the CWI
\cite{Par72}, which was derived, for gauge field theories, in the canonical
quantization \cite{Nie73} and in the path integral formulation \cite{Sar74}.
Using the latter approach the  CWI needed for conformal composite operators
was written down in the dimensional regularization \cite{Mue91a,Mue94}. To
have a convenient form for the Ward identities, we introduce a few shorthand
notations. The symbol $[O]$ means renormalization of the operator $O$ in the
$\overline{\rm MS}$-scheme. For simplicity we assume that the composite
operators are closed under renormalization. Because of Poincar\'e invariance
the $Z$-matrix is triangular (a detailed discussion on this point can be
found, for instance, in \cite{BroDamFriLep86}):
\begin{eqnarray}
\label{renOp}
[O_{kl}]=	\sum_{k'=0}^{k} Z_{kk'} O_{k'l}.
\end{eqnarray}
Furthermore, $X=\phi(x_1)\dots\phi(x_n)$ is a monomial of elementary fields
and $\langle A\rangle $ denotes the vacuum expectation value of the
time-ordered product ${\rm T} A \exp i[S]$. Then the conformal Ward
identities for the renormalized composite operators $[O_{kl}](\nu_1,\nu_2)$
finally read\footnote{These CWI do not rely on any assumptions about the
conformal symmetry breaking and they differ from the ones used in a previous
study in Ref.\ \cite{FerGriPar73} by the triangularity of the anomalous
matrices.}:
\begin{eqnarray}
\label{CWI-confOpe}
i{\cal D}\langle [O_{kl}] X \rangle &=&
	\sum_{k'=0}^{k}\left[ (l+d_1^{\rm can}+d_2^{\rm can})\delta_{kk'}+
						\gamma_{kk'}\right] \langle [O_{k'l}] X \rangle +
	\frac{\beta}{g} \langle [O_{kl}\Delta^\beta] X \rangle + \cdots,
\nonumber\\
i{\cal K}_-\langle [O_{kl}] X \rangle &=&
	-i\sum_{k'=0}^{k} \left[
	2(k-l)(k+l+\nu_1+\nu_2)\delta_{kk'}+\gamma^c_{kk'}(l)\right]
								\langle [O_{k'l-1}] X \rangle +
\\
&&\hspace{7.8cm}\frac{\beta}{g} \langle [O_{kl}\Delta^\beta_- ] X \rangle +
\cdots.
\nonumber
\end{eqnarray}
Here ${\cal D}$ and ${\cal K}$ are differential operators, which act
on each field in the monomial $X$ as in Eq.\ (\ref{PoincareGener}),
e.g.:
\begin{eqnarray}
\label{CWI-examp}
i{\cal D}\langle [O_{kl}] \phi_1(x) \phi_2(y)\rangle =
	-(d_1+d_2+x\partial_x + y\partial_y)
				\langle [O_{kl}] \phi_1(x) \phi_2(y)\rangle,
\end{eqnarray}
where the scale dimensions are shifted by the anomalous dimensions of the
field, $d_i=d^{\rm can}_i+\gamma_i$. The operators $\Delta^\beta$ and
$\Delta^\beta_- = \tilde{n}^{\star\mu} \Delta^\beta_\mu $ arise from the
conformal symmetry breaking in the action due to the renormalization of the
coupling constant. Actually, they are given by the trace anomaly of the
energy--momentum tensor \cite{Nie77,AdlColDun77,ColDunJog77,Min76} and will
be written here in terms of the renormalized Lagrangian ${\cal L}(x)$
\cite{Nie73,Sar74}:
\begin{eqnarray}
\label{conf-Delta}
{}[\Delta^\beta] =
	i\int d^nx\, g{\partial\over\partial g} {\cal L}(x),
\quad
{}[\Delta_\lambda^\beta] =
	i\int d^nx\, 2x_\lambda\, g{\partial\over\partial g}{\cal L}(x). 
\end{eqnarray}
The ellipses in the CWI denote Green functions with operator insertions
caused by the gauge-fixing and ghost terms. Such contributions should be
absent in physical matrix elements. The expression $\gamma_{kk'}$ is the
anomalous-dimension matrix of the operators and $\gamma^c_{kk'}(l)$ denotes
the special-conformal anomaly matrix, which breaks the covariance of the
operators under infinitesimal special conformal transformations. Such
transformations break the Poincar\'e invariance ($K_\alpha$ does not
commute with $M_{\beta\gamma}$ and $P_{\beta}$) and therefore the
spin $l$ dependence appears.

\subsection{Leading order analysis}

It is well known that the anomalous-dimension matrix of conformal
two-particle operators is diagonal to LO:
$\gamma_{kk'}^{(0)}=\gamma_k^{(0)} \delta_{kk'}$. However, in a general
renormalization scheme the irreducible conformal representation is already
broken by an off-diagonal special-conformal anomaly matrix. Using the
dimensional regularization and the $\overline{\rm MS}$ prescription the
results for the $\phi^3_{(6)}$ theory read, in matrix notation \cite{Mue91a}:
\begin{eqnarray}
\label{CWI-scBre1}
\hat{\gamma}^{c(0)}(l)&=&-\hat{b}(l)\hat{\gamma}^{(0)}, \qquad
\hat{b}(l)=\left\{b_{kk'}(l)\right\}, \quad
\hat{\gamma}^{(0)}=\left\{\gamma_k^{(0)} \delta_{kk'}\right\},
\\
&&\hspace{-4cm}{\mbox{\ where}} 
\nonumber\\
\label{def-b}
b_{kk'}(l)&=&
\left\{
{
   2(l+k'+3)\delta_{kk'} - 2(2k'+3) \quad
                          \mbox{if } k-k'\ge 0 \mbox{ and even}
\atop 
                             0 \hspace{3.2cm} \mbox{otherwise.}
}
\right. 
\end{eqnarray}
Regularization of the ``gluon" propagator via Pauli--Villars provides a 
different breaking of the special conformal transformation:
\begin{eqnarray}
\label{CWI-scBre1-PV}
\hat{\gamma}^{c(0)}(l)&=&-\hat{\gamma}^{(0)}\hat{b}(l).
\end{eqnarray}
However, this breaking can  easily be absorbed in a redefinition of the
local operator by changing the index\footnote{
Here $\alpha_s$ denotes of course an appropriate definition of the coupling
constant in the scalar theory. Note that only at one-loop order the
conformal symmetry prediction for the eigenfunctions of the evolution kernel
coincides with the shift of the index $\nu$ for the operator.
}
$\nu \to \nu - \frac{\alpha_s}{4\pi} \gamma_k^{(0)}$ and it coincides 
with the prediction of conformal symmetry \cite{BroDamFriLep86}.

For the conformal flavour non-singlet quark operators in QCD
the symmetry breaking is even more complicated \cite{Mue94} due to the
covariant derivatives:
\begin{eqnarray}
\label{CWI-scBre2}
\hat{\gamma}^{c(0)}(l)&=& -\hat{b}(l) \hat{\gamma}^{(0)}+\hat{w},
\end{eqnarray}
where the non-vanishing elements of $w_{kk'}$ (for $k-k'>0$ and $k-k'$
even) are
\begin{eqnarray}
w_{kk'}&=&-4C_F(2k'+3)(k-k')(k+k'+3)
\nonumber\\
&& \times 
 \left[{A_{kk'}-\psi(k+1)+\psi(0)\over (k'+1)(k'+2)}+
        {2A_{kk'}\over (k-k')(k+ k'+3)}\right] ,
\\
A_{kk'}&=&
		 \psi\left({k+k'+4\over 2}\right)-\psi\left({k-k'\over 2}\right)
       +2\psi\left(k-k'\right)-\psi\left(k+2\right)-\psi(1), 
\nonumber
\end{eqnarray}
with $\psi\left(z\right)= {d\over dz} \ln\Gamma(z)$. It turns out that the
appearance of the matrix $\hat{w}$ explains the difference between the
conformal symmetry prediction for the eigenfunctions of the pion evolution
kernel and the explicit NLO calculation.

Normalization conditions, which are given implicitly in the $\overline{\rm
MS}$ scheme, are a matter of convenience, and changing them does not affect
physical quantities. Thus we can look for a scheme in which conformal
covariance is restored. Such a scheme can be obtained by a finite
renormalization; at LO we define the renormalized conformally covariant
operators as:
\begin{eqnarray}
O_{kl}^{\rm co}&=&
	[O_{kl}] - \frac{\alpha_s}{2\pi}
		\sum_{k'=0}^{k-2}
	\frac{\gamma^{c(0)}_{kk'}(l)}{2(k'-k)(k'+k+\nu_1+\nu_2)}
	[O_{k'l}] + \dots,
\\
&=&
	O_{kl} +\frac{\alpha_s}{2\pi} \frac{\gamma^{(0)}_k}{2\epsilon} O_{kl} -
     \frac{\alpha_s}{2\pi}
		\sum_{k'=0}^{k-2}
	\frac{\gamma^{c(0)}_{kk'}(l)}{2(k'-k)(k'+k+\nu_1+\nu_2)}
	O_{k'l} + \dots,
\nonumber
\end{eqnarray}
where $1/\epsilon$ is the usual $\epsilon$-pole in dimensional regularization,
which fulfil the CWI:
\begin{eqnarray}\
\label{CWI-confOpe-LO-true}
i{\cal D}\langle O_{kl}^{\rm co} X \rangle &=&
	\left[
	l+d_1^{\rm can}+d_2^{\rm can}+\frac{\alpha_s}{2\pi}\gamma^{(0)}_{k}
	\right]
		\langle O_{kl}^{\rm co} X \rangle + \cdots,
\\
i{\cal K}_-\langle O_{kl}^{\rm co} X \rangle &=&
	-i\left[
	2(k-l)
(k+l+\nu_1+\nu_2)+ \frac{\alpha_s}{2\pi} \gamma^{c(0)}_{kk}(l)\right]
	\langle O_{kl-1}^{\rm co} X \rangle + \cdots.
\nonumber
\end{eqnarray}
For the scalar theory as well as for  QCD we find from 
Eqs.\ (\ref{CWI-scBre1}) -- (\ref{CWI-scBre2}) that 
$\gamma^{c(0)}_{kk}(l)=2(k-l)\gamma^{(0)}_k$. Thus, in both equations
of the CWI (\ref{CWI-confOpe-LO-true}) the conformal symmetry
{\em breaking} by the anomalous dimension is {\em absorbed} into the shift
of the canonical dimension of the operators
$l+d_1^{\rm can}+d_2^{\rm can} \to l+d_1^{\rm can}+d_2^{\rm can}+\gamma_k$.
In this way the irreducible conformal representation is {\em restored} for the
renormalized operator at LO. So as conformal covariance at
tree level is sufficient for a diagonal anomalous-dimension matrix at
LO, the one-loop renormalized conformal operators do not mix
under renormalization in NLO, of course, up to a term proportional to
$\beta_0$.

\subsection{Restoration of conformal covariance}

Before we use conformal constraints to extend the analysis to the full
conformal theory, let us show that the property
$\gamma^{c(0)}_{kk}(k)=0$ holds true generally. To make
the discussion transparent, let us first consider the scalar theory
in which $\hat{\gamma}^c(l)$ is defined in the $\overline{\rm MS}$ scheme as
\cite{Mue91a}
\begin{eqnarray}
\label{def-gammac-sca}
\hat{\gamma}^c(l)=
	-2\gamma_\varphi \hat{b}(l) + 2 [\hat{Z}^{[1]},\hat{b}(l)] +
	\hat{Z}^{\star[1]}(l),
\end{eqnarray}
where the counterterm
	$\hat{Z}^{\star}=\hat{Z}^{\star[1]}/\epsilon+
\hat{Z}^{\star[2]}/\epsilon^2+\dots$
has to be computed from the renormalization of the operator product
\begin{eqnarray}
\label{def-gammac-ren}
[O_{kl}][\Delta_-^{\beta}]=[O_{kl}\Delta_-^{\beta}] +
						i \sum_{k'=0}^{k} Z^\star_{kk'}(l) [O_{k'l}].
\end{eqnarray}
From the properties of $\hat{b}(l)$ and $\hat{Z}$ it follows that
$\gamma^{c}_{kk}(k)=Z^{\star[1]}_{kk}(k)$. The latter is determined
by the UV-divergent part of
	$\int d^nx\, (\tilde{n}^\star x) \varphi^3(x)\, O_{kk}$.
Since $O_{kk}$ is a polynomial of order $k$ in the derivatives
and the UV divergence is concentrated in $x^\alpha=0$, it is clear that
$\tilde{n}^\star x$ annihilates one derivative, so that 
a polynomial of order $k-1$ remains. Thus, no counterterm $O_{kk}$ is
needed and therefore $\gamma^{c}_{kk}(k)=\hat{Z}^{\star[1]}_{kk}(k)=0$.
In gauge field theories the definition of $\hat{Z}^{\star}$ is modified by
a term containing the functional derivative with respect to the gauge
field \cite{Mue94}:
$\int d^nx\, (\tilde{n}^\ast  x) A_{\mu} 
                          {\delta \over \delta A_\mu} [O_{kk}]$.
Obviously, we can use the same arguments as above, and 
this term also does not induce a contribution to $\gamma^{c}_{kk}(k)$.

The proof that conformal covariance can be restored in the conformal limit
will be achieved in the following manner. First we notice that the
anomalous-dimension matrix will be diagonalized by a finite renormalization
group transformation,
\begin{eqnarray}
\label{CWI-confOpe-true-dila}
i{\cal D}\langle O_{kl}^{\rm co} X \rangle &=&
	 \left(l+d_1^{\rm can}+d_2^{\rm can}+\gamma_k\right)
\langle O_{kl}^{\rm co} X \rangle + \cdots.
\end{eqnarray}
Then we show, with the help of conformal constraints, that this implies 
\begin{eqnarray}
\label{CWI-confOpe-true-conf}
i{\cal K}_-\langle O_{kl}^{\rm co} X \rangle &=&
	-2i(k-l)[k+l+\nu_1+\nu_2 +\gamma_k]
	\langle O_{kl-1}^{\rm co} X \rangle + \cdots.
\end{eqnarray}
To calculate the r.h.s.\ of Eq.\ (\ref{CWI-confOpe-true-conf}) we solve the
conformal constraints in the following Subsection. In a second
Subsection we consider the flavour-singlet channel in QCD, where an
additional mixing problem between quark and gluon operators appears.
Here, we only take into account the algebraic properties of the 
constraints and the triangularity of the matrices to show that
Eq.\ (\ref{CWI-confOpe-true-dila}) implies
Eq. (\ref{CWI-confOpe-true-conf}).

\subsubsection{Solution of the conformal constraints}

A constraint for the anomalous-dimension matrix,
which allows the off-diagonal part  to be computed in terms of the
special-conformal anomaly matrix, is implied by the commutator relation
\begin{eqnarray}
\label{conCon-1}
[{\cal D},{\cal K}_-] = i {\cal K}_-.
\end{eqnarray}
Applying this identity to the Green functions and using the CWI provides
immediately a commutator relation for the anomalous-dimension matrix and the
special-conformal anomaly matrix. For completeness, we give the exact
result, which includes the full $\beta$-dependence \cite{Mue91a,Mue94} in
the dimensional regularization for the $\overline{\rm MS}$
prescription\footnote{The derivation is tricky
and all details for the scalar theory in dimensional regularization are
given in \cite{Mue91a}. The calculation for the Abelian gauge field theory is
analogous and as expected leads to no explicit gauge dependence in the
commutator relation. Since $\hat{\gamma}$ is a physical quantity we can
assume that, at least for $\beta=0$, this constraint holds also true in QCD.}:
\begin{eqnarray}
\label{conf-constr-KD}
\left[\hat{a}(l)+\hat{\gamma}^c(l)
+2{\beta\over g}\hat{b}(l)
,\hat{\gamma}\right]=0, \quad
a_{kk'}(l)= 2(k-l)(k+l+\nu_1+\nu_2) \delta_{kk'}.
\end{eqnarray}
Since the matrix $\hat{a}$ is diagonal, a recurrence relation follows for
the off-diagonal part $\hat{\gamma}^{ND}$ of the anomalous-dimension matrix
($\beta$ is now consequently set to zero):
\begin{eqnarray}
\label{conf-constr-KD-1}
\gamma^{ND}_{kk'}=
		- \left\{{\cal G}\hat{\gamma}^{D}\right\}_{kk'}
		- \left\{{\cal G}\hat{\gamma}^{ND}\right\}_{kk'}.
\end{eqnarray}
Here $\hat{\gamma}^{D}=\left\{\gamma_k \delta_{kk'}\right\}$ denotes the
diagonal part of the anomalous-dimension matrix and the operator ${\cal G}$ is
defined by
\begin{eqnarray}
\label{adcodtb}
{\cal G}\hat{A}:= \left\{ { 
     {\left[\hat{\gamma}^c(l),\hat{A}\right]_{kk'}
                           \over 2(k-k')(k+k'+\nu_1+\nu_2)}                    
                           \quad
                            \mbox{if } k-k'>0 \atop 
    \hspace{1.2cm} 0   \hspace{1.7cm}    \mbox{otherwise.} 
                          }\right.
\end{eqnarray}
The solution of Eq.\ (\ref{conf-constr-KD-1}) can be formally written as
\begin{eqnarray}
\label{adcodt}
\hat{\gamma}^{ND} = - {{\cal G} \over 
                          \hat{1} + {\cal G}} \hat{\gamma}^{D}
= -{\cal G} \hat{\gamma}^{D} + 
       {\cal G}^2 \hat{\gamma}^{D}  -  \cdots.
\end{eqnarray}

The composite operators, which do not mix under renormalization, are
obtained by a finite renormalization 
\begin{eqnarray}
\label{transformation}
O_{kl}^{\rm co}=\sum_{k'=0}^{k} B^{-1}_{kk'} [O_{k'l}].
\end{eqnarray}
The matrix $\hat{B}=\{B_{kk'}\}$ can be calculated from
\begin{eqnarray}
\label{transformation-1}
\hat{\gamma}^{D} = \hat{B}^{-1}\hat{\gamma}\hat{B}
\Longrightarrow
\left[\hat{B},\hat{\gamma}^{D}\right]=\hat{\gamma}^{\rm ND}\hat{B},
\end{eqnarray}
where the diagonal matrix $\hat{\gamma}^{D}$ consists of the eigenvalues of
the triangular anomalous-dimension matrix. The solution
of this equation is
\begin{eqnarray}
\label{sol-transformation-1}
\hat{B}=
	\frac{\hat{1}}{\hat{1}-{\cal L}\hat{\gamma}^{\rm ND}}
	=\hat{1}+{\cal L}\hat{\gamma}^{\rm ND}+
	 {\cal L}\left(\hat{\gamma}^{\rm ND}{\cal L}\hat{\gamma}^{\rm
	ND}\right)+\cdots,
\end{eqnarray}
where the operator ${\cal L}$ is defined by
\begin{eqnarray}
\label{def-L}
{\cal L}\hat{A}:=
\left\{{   -{A_{kk'}\over \gamma_k-\gamma_{k'}}\quad        
                     \mbox{if } k-k' >0 \atop
       \hspace{0.3cm}   0     \hspace{0.9cm}   \mbox{otherwise.} 
                          }\right.
\end{eqnarray}
The off-diagonal matrix
$\hat{\gamma}^{\rm ND}$ is given in terms of
$\hat{\gamma}^{\rm c}$, implying that the transformation matrix
$\hat{B}$ can  also be expressed by the special-conformal anomaly matrix.
From Eqs.\ (\ref{adcodt}) and (\ref{sol-transformation-1}) one finds,
after some algebra, that the diagonal anomalous-dimension matrix cancels
out (see Appendix A):
 \begin{eqnarray}
\label{blefdt-1}
\hat{B} = {\hat{1} \over \hat{1}+{\cal J}\hat{\gamma}^c}
   = \hat{1}-{\cal J}\hat{\gamma}^c + {\cal J}(\hat{\gamma}^c 
            {\cal J}\hat{\gamma}^c)-\cdots,
\end{eqnarray}
where the operator ${\cal J}$ is defined by
\begin{eqnarray}
{\cal J}\hat{A}:=
\left\{{     {A_{kk'}\over 2(k-k')(k+k'+\nu_1+\nu_2)}\quad    
                         \mbox{if } k-k' >0 \atop
          \hspace{1cm}       0     \hspace{1.9cm}   \mbox{otherwise.} 
                          }\right.
\end{eqnarray}

Note that in the forward case all operators $O_{kl}$ with $l>k$ vanish. 
Thus, the renormalization group transformation (\ref{transformation}) does
not affect the minimal subtraction prescription in the forward case:
$O_{kk}^{\rm co}=[O_{kk}]$.

Now we are almost able to prove that the operators $O_{kl}^{\rm co}$ are
conformally covariant. For this purpose we need the spin dependence of the
special-conformal anomaly matrix, which is constrained by the
commutator relation:
\begin{eqnarray}
\label{conCon-2}
[{\cal K_-},{\cal P}_+] =	-2i ({\cal D}+{\cal M}_{-+}). 
\end{eqnarray}
Applying this relation to the Green functions and using the
CWI (\ref{CWI-confOpe}) provides
\begin{eqnarray}
\label{conf-constr-KP}
\hat{\gamma}^c(l+1)-\hat{\gamma}^c(l)=-2\hat{\gamma}.
\end{eqnarray}
The solution of this recurrence relation gives the spin dependence
of the special-conformal anomaly matrix 
\begin{eqnarray}
\label{sol-conf-constr-KP}
\gamma^c_{kk'}(l)= \gamma^c_{kk'}(k) + 2(k-l) \gamma_{kk'},
\mbox{\ and for\ } k=k':\mbox{\ }
\gamma^c_{kk}(l)= 2(k-l) \gamma_{k},
\end{eqnarray}
where the last equation follows from the above shown property
$\gamma^c_{kk}(k)=0$.
The special conformal anomaly of the 
operators $ O_{kl}^{\rm co}$ are given by
$\hat{B}^{-1}\left[\hat{a}(l)+\hat{\gamma}^c(l)\right]\hat{B}$
and using the corresponding definitions and the property
(\ref{sol-conf-constr-KP}),
a straightforward calculation given in the Appendix A provides 
\begin{eqnarray}
\label{conAno-for-coO}
\left\{
	\hat{B}^{-1}
\left[\hat{a}(l)+\hat{\gamma}^c(l)\right]\hat{B}
\right\}_{kk'}=
	2(k-l)(k+l+\nu_1+\nu_2+\gamma_k)\delta_{kk'},
\end{eqnarray}
which is equivalent to (\ref{CWI-confOpe-true-conf}).

\subsubsection{Additional mixing problem in the QCD singlet channel}

The leading twist operators appearing in the singlet channel
can be written in the following conformally covariant manner,
where for even parity we have:
\begin{eqnarray}
{^q\!O}_{kl}&=&
	\partial^l_+ \bar{\psi}\; \tilde{n}_\alpha \gamma^\alpha C^{\frac{3}{2}}_k
	\left(\frac{\tilde{n}D_-}{\tilde{n}\partial_+}\right) \psi,
\quad l\ge k\ge 0,
\\
{^g\!O}_{kl}&=&
	\partial^{l-1}_+ \tilde{n}^\alpha F_{\alpha\beta} C^{\frac{5}{2}}_{k-1}
	\left(\frac{\tilde{n}D_-}{\tilde{n}\partial_+}\right)
	F^{\beta\gamma}\tilde{n}_\gamma, \quad l\ge k\ge 1,
\end{eqnarray}
while for odd parity:
\begin{eqnarray}
{^q\!R}_{kl}&=&
	\partial^l_+ \bar{\psi}\; \tilde{n}_\alpha \gamma^\alpha\gamma^5
	C^{\frac{3}{2}}_k
	\left(\frac{\tilde{n}D_-}{\tilde{n}\partial_+}\right) \psi,
\quad l\ge k\ge 0,
\\
{^g\!R}_{kl}&=&
	\partial^{l-1}_+ \tilde{n}^\alpha \tilde{F}_{\alpha\beta}
	C^{\frac{5}{2}}_{k-1}
	\left(\frac{\tilde{n}D_-}{\tilde{n}\partial_+}\right)
	F^{\beta\gamma}\tilde{n}_\gamma,
\quad l\ge k\ge 1.
\end{eqnarray}
Here flavour and colour indices are suppressed for simplicity, 
$D_-^\nu=\stackrel{\rightarrow}{D^\nu} - \stackrel{\leftarrow}{D^\nu}$ 
are the covariant derivatives, $F_{\alpha\beta}$ and
$\tilde{F}_{\alpha\beta}=\epsilon_{\alpha\beta\gamma\delta}
F^{\gamma\delta}/2$ are the field strength and the dual-field strength
tensor, respectively. Since the dimension of the field strength tensor
is 2 and its spin is 1 the index of the Gegenbauer polynomials for the
gluon operators is $\nu=5/2$. All these operators have spin $l+1$ and
canonical dimension $l+3$, as well as the {\em same} behaviour
under special conformal transformations at tree level.

The following discussion is valid for operators of even and odd parity.
The quark and gluon operators will mix, and the
anomalous-dimension matrix of the operators can therefore be written in the
following compact notation:
\begin{eqnarray}
\hat{\gamma}=
	\left({
		{^{qq}\!\hat{\gamma}}\;{^{qg}\!\hat{\gamma}}\atop
		{^{gq}\!\hat{\gamma}}\;{^{gg}\!\hat{\gamma}}
	}\right),
\end{eqnarray}
where the entries ${^{ij}\!\hat{\gamma}}$ for $i,j=\{q,g\}$ are triangular
matrices. At LO these entries are diagonal and the remaining
mixing problem has to be solved by explicit diagonalization of the
$2\times2$ matrix.

With the previous assumption about the unphysical part in the CWI,
the generalization of the conformal constraint (\ref{conf-constr-KD})
for $\beta=0$ is a purely algebraic task: 
\begin{eqnarray}
\label{conf-constr-KD-Sing}
\left[\left({\hat{a}(l)\quad 0\atop\;\;\; 0\quad \hat{a}(l)}\right)+
\left({
		{^{qq}\!\hat{\gamma}}^c(l)\;{^{qg}\!\hat{\gamma}}^c(l)\atop
		{^{gq}\!\hat{\gamma}}^c(l)\;{^{gg}\!\hat{\gamma}}^c(l)
	}\right),
\left({
		{^{qq}\!\hat{\gamma}}\;{^{qg}\!\hat{\gamma}}\atop
		{^{gq}\!\hat{\gamma}}\;{^{gg}\!\hat{\gamma}}
	}\right)
\right]=0, 
\end{eqnarray}
where $a_{kk'}(l)= 2(k-l)(k+l+3) \delta_{kk'}$.
Now we introduce the matrix
\begin{eqnarray}
\hat{B}=
		\left({
		{^{qq}\!\hat{B}}\;{^{qg}\!\hat{B}}\atop
		{^{gq}\!\hat{B}}\;{^{gg}\!\hat{B}}
	}\right), 
\end{eqnarray}
implying that the anomalous-dimension matrices of the operators
\begin{eqnarray}
\label{transformation-sin}
{^q\!O}_{kl}^{\rm co}=
	\sum_{k'=0}^{k}\left( {^{qq}\!B}^{-1}_{kk'} [{^q\!O}_{k'l}] +
					{^{qg}\!B}^{-1}_{kk'} [{^g\!O}_{k'l}]\right),
\quad
{^g\!O}_{kl}^{\rm co}=
	\sum_{k'=0}^{k} \left( {^{gq}\!B}^{-1}_{kk'} [{^q\!O}_{k'l}] +
					{^{gg}\!B}^{-1}_{kk'} [{^g\!O}_{k'l}]\right)
\end{eqnarray} 
consist only of diagonal entries
\begin{eqnarray}
\label{sol-transformation-1-sing}
\hat{\gamma}^{\rm D} =
	\hat{B}^{-1}\hat{\gamma}\hat{B}
	=
\left({
		{^{qq}\!\hat{\gamma}}^{\rm D}\;{^{qg}\!\hat{\gamma}}^{\rm D}\atop
		{^{gq}\!\hat{\gamma}}^{\rm D}\;{^{gg}\!\hat{\gamma}}^{\rm D}
	}\right),
\end{eqnarray}
where ${^{ij}\!\gamma}^{\rm D}_{kk'}={^{ij}\!\gamma}_k \delta_{kk'}.$
Applying the transformation (\ref{sol-transformation-1-sing}) to 
the conformal constraint (\ref{conf-constr-KD-Sing}) tells us
that 
\begin{eqnarray}
\hat{R}(l)=
\hat{B}^{-1}\left[
\left({\hat{a}(l)\quad 0\atop\;\;\; 0\quad \hat{a}(l)}\right)+
\left({
		{^{qq}\!\hat{\gamma}}^c(l)\;{^{qg}\!\hat{\gamma}}^c(l)\atop
		{^{gq}\!\hat{\gamma}}^c(l)\;{^{gg}\!\hat{\gamma}}^c(l)
	}\right)\right]
\hat{B}
\end{eqnarray}
possesses only diagonal entries. Taking into account the property
$\hat{a}(l+1)-\hat{a}(l)=-2(l+3)\hat{1}$ and the analogous
equation to Eq.\ (\ref{conf-constr-KP}) it follows that
\begin{eqnarray}
\label{conf-constr-KP-sin}
\hat{R}(l+1)-\hat{R}(l) = -2(l+3)
\left({
		\hat{1}\; 0
\atop
		0\; \hat{1}
	}\right) -2
\left({
		{^{qq}\!\hat{\gamma}^{\rm D}}\;{^{qg}\!\hat{\gamma}^{\rm D}}\atop
		{^{gq}\!\hat{\gamma}^{\rm D}}\;{^{gg}\!\hat{\gamma}^{\rm D}}
	}\right).
\end{eqnarray}
The solution of this recurrence
relation together with the property $R_{kk}(k)=0$ for
$i,j=\{q,g\}$ gives the following term in the r.h.s.\ of the special
conformal Ward identity:
\begin{eqnarray}
R_{kk'}(l)=2(k-l)\left({
	\hspace{-1cm}k+l+3+{^{qq}\!\gamma}_k
	\hspace{1cm}{^{qg}\!\gamma}_k\atop
	\hspace{1cm}{^{gq}\!\gamma}_k
	\hspace{1.2cm} k+l+3+{^{gg}\!\gamma}_k
	}\right) \delta_{kk'},
\end{eqnarray}
which shows that the lowest member of each tower will be annihilated by
the action of  $K_-$ in the conformal limit of the theory.

To restore the conformal covariance completely, 
the remaining mixing problem that is well known from the forward case
has to be solved by introducing the eigenvectors:
\begin{eqnarray}
{^+\!O}^{\rm co}_{kl}=
	{^q\!O}^{\rm co}_{kl} + C^+_k\; {^g\!O}^{\rm co}_{kl},
\quad
{^-\!O}^{\rm co}_{kl}=
	{^g\!O}^{\rm co}_{kl} + C^-_k\; {^q\!O}^{\rm co}_{kl}.
\end{eqnarray}
These operators have completely diagonal anomalous dimension
and special conformal anomaly matrices:
\begin{eqnarray}
\hat{\gamma}=
	\left({
		\!{^{+}\!\hat{\gamma}^{\rm D}}\;\;\;0\atop
		\;\;0\;\;{^{-}\!\hat{\gamma}^{\rm D}}
	}\right),
\quad
\hat{\gamma}^{\rm c}=
	2(k-l) \left({
		\!{^{+}\!\hat{\gamma}^{\rm D}}\;\;\;0\atop
		\;\;0\;\;{^{-}\!\hat{\gamma}^{\rm D}}
	}\right).
\end{eqnarray}

\subsection{Conformally covariant OPE in the interacting theory}
\label{ConRenSch-inttheo}

In the previous Subsection it has been proved that there exists, in the
hypothetical conformal limit of the theory, a renormalization scheme in which
the conformal covariance of the renormalized operators is ensured. This scheme
is related to any other one by a finite renormalization of the composite
operators, where, however, in the forward case the normalization of these
operators remains unchanged. In the OPE the renormalized operators are
normalized at the factorization scale. These normalization conditions are
arbitrary and the physical quantities, which are defined in terms of the
product of two currents, are independent of these conditions. Now we choose
such normalization conditions that ensure the covariance of the renormalized
conformal operators.

Because of the covariance, the operators in different towers do not mix
under dilatations, and the dilatation
invariance requires the form of the OPE given in Eq.\ (\ref{OPE-def}), where
the scaling dimension of the renormalized currents and of the renormalized
conformal operators is now $l_i=l_i^{\rm can}+\gamma_i$ for $i=\{A,B\}$ and
$l_k=d_1^{\rm can}+d_2^{\rm can}+k+\gamma_k$, respectively. Furthermore,
the renormalized conformal operators transform under infinitesimal special
conformal transformations formally as in Eq.\ (\ref{cOPE-act-K}), so that the
lowest member of each conformal tower will be annihilated by $K_{\lambda}$.
Hence, we can apply the same algebraic steps as previously; the result will be
the same formal expression  as (\ref{cOPE-def}) also for the interacting
theory, however, with shifted scale dimensions:
\begin{eqnarray}
\label{cOPE-def-int}
A(x) B(0) &=&
	\sum_{k=0}^\infty
	\tilde{C}_k(\mu) \left(\frac{1}{x^2}\right)^{
(l_A+l_B-d_1^{\rm can}-d_2^{\rm can}-\gamma_k)/2} 
	(-i)^{(k+1)} x^{\alpha_1}\dots x^{\alpha_k}\times
\\
	&&\int_0^1du\, u^{(l_A-l_B)/2} (1-u)^{(l_B-l_A)/2}
	[u(1-u)]^{(d_1^{\rm can}+d_2^{\rm can}+\gamma_k)/2+k-1}
	O^{\rm co}_{\alpha_1\dots\alpha_k}(ux)_\mu. 
\nonumber
\end{eqnarray}
In the conformal factorization scheme, the Wilson coefficients and the
composite operators satisfy simple renormalization group equations:
\begin{eqnarray}
\label{cOPE-RGE-Coe}
\mu\frac{\partial}{\partial \mu}\tilde{C}_k(\mu)&=&
	(\gamma_k-\gamma_A-\gamma_B)\tilde{C}_k(\mu),
\\
\label{cOPE-RGE}
\mu\frac{\partial}{\partial \mu} O^{\rm co}_{\alpha_1\dots\alpha_k}(0)_\mu
&=& -\gamma_k O^{\rm co}_{\alpha_1\dots\alpha_k}(0)_\mu.
\end{eqnarray}

The conformally covariant OPE for the product of two electromagnetic
currents in QCD should be constructed at leading twist-2 in an analogous
way. However, to avoid technical complications due to the Lorentz structure
and the gauge-invariant decomposition, we consider here only two independent
contributions, namely the trace $J_\mu(x)J^\mu(0)$ and the antisymmetric
twist-2 part $ \left\{J_\mu(x)J_\nu(0)\right\}^{\rm asy}$ proportional to the
$\epsilon$-tensor. The neglected twist-3 contributions induce in certain
cases non-power-suppressed contributions as for the structure function $g_2$
in polarized DIS. In both cases  considered, the Lorentz structure does not
affect the derivation of the conformal OPE, and we can proceed in principle
as in the scalar theory. Of course, the conformal OPE is separately valid
for the flavour non-singlet and singlet channel. The appearing mixing
problem in the singlet channel can be resolved easily, as discussed above,
by introducing appropriate linear combinations of quark and gluon operators,
which will be considered as independent. Hence, in the conformal limit, each
Wilson coefficient that appears can be written in the same form as given in
(\ref{cOPE-def-int}):
\begin{eqnarray}
\label{cOPE-def-QCD}
\left\{
	{J_\mu(x) J^\mu(0) \atop \left\{J_\mu(x) J_\nu(0)\right\}^{\rm asy}}
\right\}
 &=& \sum_{k=0}^\infty \sum_{i}
\left\{{\hspace{0.5cm}
	x^{\alpha_1}  \hspace{0.8cm} {^i\!\tilde{C}}_{k}(\mu)
	\atop
	i\epsilon_{\mu\nu\lambda}^{\mbox{\ \ \ \ }\alpha_1} x^{\lambda}\;
										{^i\!\tilde{E}}_{k}(\mu)
}\right\}
	 \left(\frac{1}{x^2}\right)^{2-\frac{1}{2}{^i\!\gamma}_k}\times
\\
& & (-i)^{k+1} x^{\alpha_2}\dots x^{\alpha_k+1}
	\int_0^1 du\, [u(1-u)]^{k+1+\frac{1}{2}{^i\!\gamma}_{k}}\; 
\left\{
	{{^i\!O}^{\rm co}_{\alpha_1\dots \alpha_{k+1}}(ux)  \atop
{^i\!R}^{\rm co}_{\alpha_1\dots \alpha_{k+1}}(ux)}
\right\},
\nonumber
\end{eqnarray}
where the index $i=\{\mbox{NS},+,-\}$ denotes the quark operators in the
non-singlet channel as well as the eigenvectors introduced in the singlet
channel. Note that the anomalous dimensions ${^i\!\gamma}_k$ are different
for even and odd parity operators.

To compute the Wilson coefficients one would choose, for practical purposes,
the simplest normalization conditions. Then the conformal covariance
of the OPE is not manifest and the operators will mix under renormalization.
However, putting together the solution of the more complicated
renormalization group equation and the Wilson coefficients provides the same
scheme-independent result as the conformally covariant OPE. This fact will
be used below for a consistency check of the available non-forward and
forward QCD calculations for the product of two electromagnetic currents
in NLO. \\

\section{Conformal prediction for two-photon processes \\
		in the light-cone-dominated region}
\label{Conf-Pre}

In the following we consider two-photon processes, where
at least one of the photons is far off-shell, so that the momentum transfer
is large, which means that the distance between the photons is light-like.
Such processes are the deeply virtual Compton scattering, which is
widely discussed at present \cite{Ji96,Ji97,Rad96}:
\begin{eqnarray}
\label{TP-Pro-dvCs}
  \gamma^\ast (q_1) + \hbox{H}(P_1) \to \gamma^\ast(q_2) + \hbox{H}(P_2),
\end{eqnarray}
and the production of some hadronic final states by photon--photon
collision, e.g. the crossed process to the DVCS:
\begin{eqnarray}
\label{TP-Pro-thp}
\gamma^\ast (q_1) + \gamma^\ast (q_2) \to \hbox{H}(P_1) + \hbox{H}(P_2).
\end{eqnarray}
We pay special attention to meson transition form factors $\gamma^\ast
+ \gamma^\ast \to \hbox{M}$. Note that also the production of two jets in
the light-cone-dominated region was already considered  in the beginning
of the 80's by Chase \cite{Cha80,Cha80a}.
Both processes (\ref{TP-Pro-dvCs})
and (\ref{TP-Pro-thp}) were previously studied in the framework of the
non-local light-cone expansion by Geyer, Robaschik and collaborators
at leading order \cite{GeyMueRob92,MueRobGeyDitHor94}. Here we employ 
the conformally covariant OPE to predict the leading twist-2 contributions
of these non-forward processes restricted to the conformal limit in terms
of the off-diagonal expectation values of composite operators.

\subsection{General formalism}
The scattering amplitude for two-photon processes is given by the
time-ordered product of two electromagnetic currents sandwiched between the
corresponding hadronic states.
To be more general, we define the  scattering amplitude in the momentum
space generically as
\begin{eqnarray}
\label{TP-ScaAmp}
T(P_1,P_2,q) =
   i \int d^n x\, e^{i qx}
	\langle P_2|T J\left(\frac{x}{2}\right)
                                  J\left(\frac{-x}{2}\right)|P_1\rangle,
\end{eqnarray}
where $J$ denotes a current and $n$ is again the (integer) space-time
dimension. At large momentum transfer $q=(q_1+q_2)/2$ the process is
dominated by the contributions from the light cone and we can define the
following generalized Bjorken region \cite{GeyMueRob92,MueRobGeyDitHor94}:
\begin{eqnarray}
\label{TP-gBl-1}
\nu = Pq\rightarrow \infty, \mbox{\ \ \ where\ } P=P_1+P_2, \qquad
Q^2 = -q^2 \rightarrow \infty, 
\end{eqnarray}
with the scaling variables
\begin{eqnarray}
\label{TP-gBl-2}
   \xi  =  \frac{1}{\omega}={-q^2  \over Pq} , \qquad
	\eta = {\Delta q \over Pq},
\mbox{\ \ \ where\ } \Delta=P_2-P_1.
\end{eqnarray}
In the forward case $\xi$ is identical to the Bjorken variable $x_{Bj}$,
and $\eta$ vanishes. For non-forward Compton scattering in the
Breit frame $\eta$ is  approximately given by
$\cos\phi=-(\vec{\Delta}\vec{q}/|\vec{\Delta}||\vec{q}|)$
\cite{GeyMueRob92,MueRobGeyDitHor94}. Formally, $\eta$ interpolates
between different processes, for instance the two-photon production of
one hadron requires $\eta=1$.

It is straightforward to derive the conformal predictions for the two-photon
processes in the generalized Bjorken limit by inserting the conformally
covariant OPE into the scattering amplitude (\ref{TP-ScaAmp}):
\begin{eqnarray}
\label{TP-conf-pred-st}
T(\omega,\eta,Q^2) &=&
	\int d^n x\,
		e^{i\left\{qx-\frac{1}{2}\Delta x\right\}}
		\sum_{k=0}^\infty
	\tilde{C}_k(\mu) \left(\frac{1}{x^2}\right)^{\frac{2l_J-l_k+k}{2}}
 (-i)^k x^{\alpha_1}\dots x^{\alpha_k} \times
\\
&&\qquad\qquad\qquad\qquad\int_0^1 du\; [u (1-u)]^{(l_k+k)/2-1}
\langle P_2| O^{\rm co}_{{\alpha_1}\dots {\alpha_k}}(u x)|P_1\rangle .
\nonumber
\end{eqnarray}
The expectation value
$\langle P_2| O^{\rm co}_{{\alpha_1}\dots {\alpha_k}}(u x)|P_1\rangle$
is a symmetric and traceless tensor, which can be built from the vectors
$P_{\alpha_i}$ and $\Delta_{\alpha_i}$. With respect to the Fourier
transform, where we keep only the leading
terms in $Q^2$, we already set $\Delta x= \eta Px$ as well as
$\Delta_{\alpha_i}= \eta P_{\alpha_i}$:
\begin{eqnarray}
\label{TP-conf-expec}
\langle P_2| O^{\rm co}_{\alpha_1\dots\alpha_k}(ux)|P_1\rangle_{\mu}=
P_{\alpha_1}\dots P_{\alpha_k} e^{i u\eta (Px)} 
\langle P_2| O^{\rm co}_{k}(0)|P_1\rangle_{\mu}(\eta).
\end{eqnarray}
These reduced expectation values are polynomials
of order $k$ in $\eta$ and depend on the factorization scale $\mu$.
Such off-diagonal matrix elements are universal and appear not only in
two-photon processes, but also in exclusive electroproduction of mesons
\cite{ColFraStr96,Rad96a}. They are (conformal) moments of the
off-forward parton distributions  introduced in
\cite{GeyDitHorMueRob88,Ji96,Ji97,Rad96}. Jain and Ralston
pointed out that in QCD the first moment (given by the matrix element of
a current) is related to elastic form factors \cite{JaiRal93}. A first
non-perturbative calculation of the off-forward parton distributions
in the bag model has been done recently \cite{JiMelSon97}.

Inserting the reduced expectation values (\ref{TP-conf-expec}) into
Eq.\ (\ref{TP-conf-pred-st}) provides:
\begin{eqnarray}
\label{TP-conf-pred-1}
T(\omega,\eta,Q^2) &=&
	\int_0^1 du\int d^n x\,
		e^{i\left\{qx+\eta\left(u-\frac{1}{2}\right) Px\right\}}
		\sum_{k=0}^\infty
	\tilde{C}_k \left(\frac{1}{x^2}\right)^{\frac{2l_J-l_k+k}{2}}
 \times
\\	&&\qquad\qquad  (-i xP)^k [u (1-u)]^{(l_k+k)/2-1}
\langle P_2| O^{\rm co}_{k}(0)|P_1\rangle(\eta).
\nonumber
\end{eqnarray}
Employing the representation
\begin{eqnarray}
	e^{i\left\{qx+\eta\left(u-\frac{1}{2}\right) Px\right\}}
	(-i)^{k} (xP)^k=(-\eta)^{-k} \frac{d^k}{du^k}
	e^{i\left\{qx+\eta\left(u-\frac{1}{2}\right) Px\right\}}
\end{eqnarray}
and after a Fourier transformation, using the definition of the hypergeometric
functions
\begin{eqnarray}
\label{Hyp-Def}
{_2F}_1\left({\alpha,\beta\atop \gamma}\Big|x\right)=
\frac{1}{B(\alpha,\gamma-\alpha)}
\int_0^1 du\, u^{\alpha-1} (1-u)^{\gamma-\alpha-1} (1-u x)^{-\beta},
\end{eqnarray}
the desired result reads
\begin{eqnarray}
\label{conf-T-def}
T(\omega,\eta,Q^2)=
   \sum_{k=0}^{\infty} \left(Q^2\right)^{-p}
	C_k\left(\eta\omega;Q^2/\mu^2\right)
	\left(\frac{2\omega}{1+\eta\omega}\right)^k 
\langle P_2|O^{\rm co}_k(\mu^2)|P_1\rangle(\eta),
\end{eqnarray}
where the exponent of $Q^2$ is given by
$p=(n+l_k^{\rm can}-k-2l_J^{\rm can})/2$. The $\mu$-dependence of the
coefficients $ C_k(\eta\omega;Q^2/\mu^2)$ is governed by the 
renormalization group equation (\ref{cOPE-RGE-Coe}). They are known up to
the normalization
\begin{eqnarray}
\label{conf-T-def-coe}
C_k(x;1)=c_k  (1+x)^{-\frac{n+l_k-k-2l_J}{2}}
{_{2}F}_1\left({(k+l_k)/2,(n+k+l_k-2 l_J)/2 \atop
k+l_k}\Bigg|\frac{2x}{1+x}\right),
\end{eqnarray}
where $l_k=2d^{\rm can}+k+\gamma_k$. From the properties of the
hypergeometric functions, the symmetry relation $C_k(x)=(-1)^k C_k(-x)$
follows. The overall normalization can be computed in the forward case $c_k
=C_k(0)$. The reduced expectation values $\langle P_2|O^{\rm
co}_k(\mu^2)|P_1\rangle(\eta)$ satisfy the diagonal renormalization group
equation (\ref{cOPE-RGE}), which means that the $\eta$-dependence remains
invariant under evolution and only the normalization will change.

\subsection{QCD predictions}

The conformally covariant OPE for two electromagnetic
currents can now be applied in the same manner to predict different
two-photon processes and their scattering amplitudes. In the conformal limit
the same Wilson coefficients appear for quite different processes at leading
twist-2. The process dependence comes from the non-perturbative expectation
values of the conformal operators. For instance, the coefficient ($E_k^{i}$)
 $C_k^{i}$ appears in both polarized DIS and the two-photon production of
(pseudo) scalar mesons at large momentum transfer as well as in the
kinematical decomposition of the hadronic tensor for DVCS or for hadron
production. Employing the conformal OPE (\ref{cOPE-def-QCD}) and performing
steps analogous to those in the previous Subsection leads to the prediction
for the trace and the antisymmetric twist-2 part (proportional to the
$\epsilon$-tensor) of the different hadronic tensors in the conformal
limit, which is, up to trivial kinematical factors, given by:
\begin{eqnarray}
\label{conf-T-def-QCD}
\left\{{
F(\omega,\eta,Q^2)
\atop
g_1(\omega,\eta,Q^2)
}\right\}=
   \sum_{k=0}^{\infty} \sum_{i}
\left\{{
	{^i\!C}_{k}\left(\eta\omega;Q^2/\mu^2\right)
\atop
	{^i\!E}_{k}\left(\eta\omega;Q^2/\mu^2\right)
	}\right\} \left(\frac{2\omega}{1+\eta\omega}\right)^k
\left\{{
	\langle P_2|{^i\!O}^{\rm co}_k(\mu^2)|P_1\rangle(\eta)
\atop
	\langle P_2|{^i\!R}^{\rm co}_k(\mu^2)|P_1\rangle(\eta)
}\right\},
\end{eqnarray}
where the coefficient functions read
\begin{eqnarray}
\label{conf-T-def-coe-QCD}
\left\{{
	{^i\!C}_{k}(x;1)
\atop
	{^i\!E}_k(x;1)
}\right\}
=
\left\{{ {^i\!c}_k
\atop
{^i\!e}_k
}\right\}
	(1+x)^{-1-\frac{1}{2}{^i\!\gamma}_k}
	{_{2}F}_1\left(
				{k+1+\frac{1}{2}{^i\!\gamma}_k,
				k+2+\frac{1}{2}{^i\!\gamma}_k \atop
				2k+4+{^i\!\gamma}_k}\Bigg|\frac{2x}{1+x}
\right).
\end{eqnarray}
Without restrictions the obtained predictions are valid at LO. Thus, also
the evolution equations for the off-diagonal parton distribution amplitudes
are solved easily in terms of the conformal moments \cite{Cha80a}. In NLO
the correction due to the $\beta$-function appears originally only in the
off-diagonal matrix elements of the anomalous-dimension matrices. From the
NLO calculation of the pion evolution kernel \cite{DitRad84,Sar84,Kat85,MikRad85}
and also from the conformal constraints (\ref{conf-constr-KD}), this
correction is known for the non-singlet channel. Beyond this order it is
expected that also the coefficient functions will be off-diagonal, because of
the conformal symmetry breaking proportional to the $\beta$-function.

In the forward case, i.e. $\eta=0$, the conformal expansion is related
to the moments in DIS by a dispersion relation:
\begin{eqnarray}
\label{conf-T-def-QCD-fwca}
\left\{{
F(\omega,Q^2)
\atop
g_1(\omega,Q^2)
}\right\}=
   \sum_{k=0}^{\infty} \sum_{i} \omega^k
\left\{{
	{^i\!c}_{k} \langle P|{^i\!O}_k(\mu^2)|P\rangle
\atop
	{^i\!e}_{k} \langle P|{^i\!R}_k(\mu^2)|P\rangle
}\right\},
\end{eqnarray}
where the additional factor $2^k$ has been absorbed into the reduced expectation
values [compare with (\ref{TP-conf-expec})]. Thus, the overall normalization
${^i\!C}_{k}(0;1)={^i\!c}_k$ and ${^i\!E}_{k}(0;1)={^i\!e}_k$ can be taken
from the calculations in unpolarized \cite{NeeZij91,ZijNee91a,ZijNee92}
 and polarized \cite{ZijNee94} DIS, respectively, and are known up to 
order $\alpha_s^2$.

A further special case is $\eta=1$, where the conformal 
expansions give, up to a kinematical prefactor, the amplitudes for the
production of pseudo-scalar and scalar mesons by virtual photons. Here
complete NLO calculations were performed in the non-singlet channel and
can now serve as a consistency check.

\section{Radiative corrections to meson transition \\ form factors}
\label{Conf-Appl}

\subsection{Transition form factors and conformally covariant OPE}

The photon-to-meson transition form factor, measured in 
$\gamma^\star(q_1) \gamma^\star(q_2)\to M(P) $, is given  at large
momentum transfer as a convolution of the hard-scattering amplitude
$T(\omega,x,\alpha_s)$ and the meson distribution amplitude (DA)
$\phi(x,Q^2)$ \cite{BroLep80},
\begin{eqnarray}
\label{TFF-Def-ForFac-gen}
\Gamma_{\alpha\beta}=
	\epsilon_{\alpha\beta\mu\nu} q_1^\mu q_2^\nu F(\omega,Q^2), \qquad
F(\omega,Q^2)=
\frac{N}{Q^2} T(\omega,x,\alpha_s(Q^2))\otimes\phi(x,Q^2).
\end{eqnarray}
The kinematical variables are defined as before by $\omega=Pq/Q^2$ and
$q=(q_1-q_2)/2$. The factor $N$ is determined by the underlying flavour
structure, e.g. for the $\pi^0$ meson $N=e_u^2-e^2_d$. The hard-scattering
amplitude is given perturbatively by
\begin{eqnarray}
\label{TFF-HarSca-Def}
T(\omega,x,\alpha_s) &=& 
	\hat{T}^{(0)}(\omega,x) + 
	\frac{\alpha_s}{2\pi} \hat{T}^{(1)}(\omega,x) + 
	O\left(\alpha_s^2\right) + \{x\to 1-x\},
\\
\label{TFF-HarSca-LO}
\hat{T}^{(0)}(\omega,x) &=& 
	\frac{1}{1+\omega[(1-x)-x]},
\\
\label{DA-EvoKer-Def}
V(x,y;\alpha_s)&=& 
	\frac{\alpha_s}{2\pi} [V^{(0)}(x,y)]_+ +
	\left(\frac{\alpha_s}{2\pi}\right)^2 [V^{(1)}(x,y)]_+ +
	O\left(\alpha_s^3\right),
\\
\label{DA-EvoKer-LO}
V^{(0)}(x,y)&=& C_F
	\theta(y-x) \frac{x}{y}\left(1+\frac{1}{y-x}\right) +
	\left\{x\to 1-x \atop y \to 1-y\right\},
\end{eqnarray}
where the $+$-prescription is defined as 
$[V(x,y)]_+ =V(x,y) -\delta(x-y) \int dz V(z,y)$.

The given formulas coincide at leading order with the prediction of the
conformally covariant  OPE (\ref{conf-T-def-QCD}) for
$g_1(\omega,\eta=1,Q^2)$. To make this
correspondence explicit, we expand the distribution amplitude in terms of
the eigenfunctions of the evolution kernel, which is actually given by the
conformal spin expansion
\begin{eqnarray}
\label{DA-ConSpin-Exp}
\phi(x,Q^2) = \sum_{k=0}^{\infty} {(1-x)x \over N_k} C_k^{3\over 2}(2x-1)  
\langle P| O_{kk}(\mu^2)|0\rangle^{red}_{|\mu^2=Q^2}, \quad
N_k=\frac{(k+1)(k+2)}{4(2k+3)}.
\end{eqnarray}
Taking into account the definition of the Gegenbauer polynomials 
\begin{eqnarray}
\label{Geg-Rep-1}
{(1-x)x \over N_k} C_k^{3\over 2}(2x-1)=
	(-1)^k \frac{2(2k+3)}{(k+1)!} \frac{d^k}{dx^k} [x(1-x)]^{k+1},
\end{eqnarray}
the transition form factor (\ref{TFF-Def-ForFac-gen}) reads
\begin{eqnarray}
\label{TFF-Com-cOPE-LO}
F(\omega,Q^2)&=&
	\frac{N}{Q^2}\sum_{k=0}^{\infty}\frac{2(2k+3)}{(k+1)!}  \int_0^1 dx\,
	 [x(1-x)]^{k+1}\frac{d^k}{dx^k}
	\Bigg(	\frac{1}{1+\omega[(1-x)-x]} +
\\
&&\hspace{4.5cm} \frac{1}{1+\omega[x-(1-x)]}\Bigg)
	\langle P| O_{kk}(\mu^2)|0\rangle^{red}_{|\mu^2=Q^2}.
\nonumber
\end{eqnarray}
Performing the differentiation and using the definition of hypergeometric
functions (\ref{Hyp-Def}), Eq.\ (\ref{TFF-Com-cOPE-LO}) coincides -- up to
different normalization factors for the composite operators -- with the
conformal OPE prediction (\ref{conf-T-def-QCD}) to LO:
\begin{eqnarray}
\label{TFF-Com-cOPE-LO-fin}
F(\omega,Q^2)&=&
	\frac{N}{Q^2}\sum_{k=0}^{\infty}
	B(k+1,k+2) \frac{2(2\omega)^k}{(1+\omega)^{k+1}}
{_{2}F}_1\left({k+1, k+2 \atop
2(k+2)}\Bigg|\frac{2\omega}{1+\omega}\right)\times
\\
&&\hspace{5cm} \langle P| O_{kk}(\mu^2)|0\rangle^{red}_{|\mu^2=Q^2} +
\{\omega\to-\omega\}.
\nonumber
\end{eqnarray}

\subsection{Consistency check at next-to-leading order}

Now we are able to perform the consistency check of the existing NLO
calculations. The $\alpha_s$-correction to the hard scattering part was
computed by del Aguila and Chase \cite{AguCha81} in the OPE approach and by
Braaten \cite{Bra83} (these papers contain also the corrections to the
scalar meson and the longitudinal component of the vector meson transition
form factor, respectively) as well as by Radyushkin et al.\
\cite{KadMikRad86} in the hard-scattering picture. The results are derived
in the $\overline{\rm MS}$ scheme and the occurring $\gamma^5$ ambiguity in
dimensional regularization was resolved with different methods. The results
are in agreement. To show the structure most clearly, we rewrite their
result in the following form:
\begin{eqnarray}
\label{TFF-HarSca-NLO}
\hat{T}^{(1)}(\omega,x) &=&
		\hat{T}^{(0)}(\omega,z)\otimes\Bigg[
		[V^{(0)}(z,x)]_+ \ln\frac{Q^2}{\mu^2} -
	\frac{3}{2}[V^{\rm b}(z,x)]_+ - \frac{3}{2}C_F\delta(z-x)
										\Bigg] +
\nonumber\\
					& & 
		\hat{T}^{(0)}(\omega,z)\ln\left\{1+\omega(\bar{z}-z)\right\}
					\otimes [V^{(0)}(z,x)]_+ 
		+\hat{T}^{(0)}(\omega,z)\otimes [g(z,x)]_+,
\nonumber\\
g(x,y)&=& C_F \theta(y-x) {\ln\left(1-{x\over y}\right) \over x-y}+
\left\{ {x \to 1-x \atop y \to 1-y} \right\},
\\
V^{\rm b}(x,y)&=& C_F\theta(y-x) \frac{x}{y}\frac{1}{y-x} +
	\left\{x\to 1-x \atop y \to 1-y\right\},
\nonumber
\end{eqnarray}
where only the renormalization scale was identified with the
factorization scale $\mu$, so that latter remains explicit.
The solution of the evolution equation in the conformal limit
is known and given by the conformal spin expansion
\begin{eqnarray}
\label{DA-ConSpin-Exp-BLO}
\phi(x,Q^2) = \sum_{k=0}^{\infty} \varphi_k^{\rm ef}(x,\alpha_s)
\langle P|  O_{kk}(\mu^2)|0\rangle^{red}_{|\mu^2=Q^2},
\end{eqnarray}
where the eigenfunctions of the evolution kernel can be written as
\begin{eqnarray}
\label{ef}
\varphi_k^{ef}(x,\alpha_s)\ =\ 
       (-1)^k\  {{2\left(3+2k\right)}\over\left(k+1\right)!}\ 
       {d^k\over dx^k}\left[  x^{1 + k} {\left( 1 - x \right) }^{1 + k}\  
    \left( 1+{\alpha_s \over 2\pi}\  F_k(x)+O(\alpha_s^2)\right)\right].
\end{eqnarray}
The $\alpha_s$-correction was obtained by the leading order calculation
of the special conformal anomaly matrix for parity-even operators and
employing the formula (\ref{blefdt-1}) for the transformation matrix
$\hat{B}$:
\begin{eqnarray}
\label{efcoef}
\!\!\!F_k(x)\ &=&\  {(\gamma^{(0)}_k+\beta_0)} \left[ {1\over 2}\ 
{\ln\Big( x(1-x)\Big)}
                               -{\psi}(2+k)+{\psi}(4+2k)\right] 
\nonumber\\[.2in]
&&+\ C_F \left[\ {\ln^2\left( {{\textstyle 1-x}\over \textstyle  x}\right) 
\over 2} 
\ -\   \sum_{i = 1}^{1 + k}
     \left( -{1\over i} + {1+\delta_{1i}\over {2 + k}} \right) \,
      \Big( {\phi}(1 - x,1,i) + 
        {\phi}(x,1,i)  \Big)\right. 
\nonumber\\[.2in]
&&\left.\qquad\qquad\   +\ 2\left( 
    {{\left( 3 + 2\,k \right) \,
     \Big({\psi}(2+k)-\psi(1) \Big) }\over 
{\left( 1 + k \right) \,\left( 2+k\right) }}+{\psi}'(2 + k) -
{\pi ^2\over 4} \right)\right]\ ,
\end{eqnarray}
where $\psi'(z)=d\psi(z)/dz$ and 
$\phi(x,1,i)=\sum_{k=0}^\infty x^k/(i+k)$ are the Lerch transcendent.
This result coincides with the calculated evolution kernel in NLO
\cite{DitRad84,Sar84,Kat85,MikRad85}. The authors used the naive
$\overline{\mbox{MS}}$ scheme in which $\gamma^5$ is anticommutative,
implying that the evolution kernel for pseudo-scalar mesons is the same as
for scalar ones.  For convenience we rewrite the $\alpha_s$-correction as a
convolution:   
\begin{eqnarray}
\label{EfBL-Def}
\varphi_n^{ef}(x,\alpha_s)=
\left(\delta(x-y)+{\alpha_s\over 2\pi} c^{(1)}(x,y)+\cdots\right)\otimes
 {(1-y)y\over N_n} C_n^{3\over 2}(2y-1),
\end{eqnarray}
where\footnote{The definition of $c^{(1)}(x,y)$ in \cite{Mue94,Mue95} contains
a misprint concerning the sign of $g(x,y)$.}
\begin{eqnarray}
c^{(1)}(x,y)&=&
	(I-{\cal P})\left(
	{\beta_0\over 2}S(x,y)-S(x,z)\otimes V^{(0)}(z,y)-[g(x,y)]_+
	\right).
\nonumber
\end{eqnarray}
Furthermore, the shift operator $S(x,y)$ is implicitly defined by
\begin{eqnarray}
\label{dewf-shiftop}
S(x,y)\otimes{(1-y)y\over N_n} C_n^{3\over 2}(2y-1) = 
{d\over d\rho} 
       {((1-x)x)^{1+\rho}\over N_n} C_n^{{3\over 2}+\rho}(2x-1)_{|\rho=0},
\end{eqnarray}
$I$ is the identity and the operator ${\cal P}$ projects onto the
diagonal part of the expansion of a function $f(x,y)$ with respect to
$C_i^{3/2}$; i.e.
${\cal P} f(x,y)=\sum_{i=0}^{\infty} (1-x)x/N_i C_i^{3/2}(2x-1) f_{ii}
C_i^{3/2}(2y-1)$,
where $f_{ij}$ with $0\le i,j\le\infty$ are the expansion coefficients.

The hard-scattering part and the evolution kernel were computed in
the same scheme. The convolution of the hard-scattering part
(\ref{TFF-HarSca-NLO}) with the solution  (\ref{EfBL-Def}) for
$\beta =0$ yields:
\begin{eqnarray}
\label{TFF-Def-ForFac-NLO}
F&=&
	\frac{N}{Q^2}\sum_{k=0}^{\infty}
	\left(	T^{(0)} + \frac{\alpha_s}{2\pi}{\cal T}^{(1)} 
		\right)\otimes	{(1-x)x \over N_k} C_k^{3\over 2}(2x-1)
 \langle P|  O_{kk}|0\rangle^{red}.
\end{eqnarray}
The off-diagonal part of $g(x,y)$ in the hard-scattering amplitude and
the eigenfunctions cancel with each other and  only the diagonal part is
left. After decomposition of the term 
$\hat{T}^{(0)}(\omega,z)\ln\left\{1+\omega(\bar{z}-z)\right\}$
into a diagonal and an off-diagonal part\footnote{
Diagonal refers to terms that contribute only to the normalization of
the partial waves in the expansion (\ref{TFF-Com-cOPE-LO-fin}), while
off-diagonal terms cannot be represented in such an expansion with the
given hypergeometric functions.}
the $\alpha_s$ correction to the hard-scattering amplitude reads
symbolically
\begin{eqnarray}
\label{TFF-HarSca-NLO-mod}
\hat{{\cal T}}^{(1)}(\omega,x)&=&
\hat{T}^{(0)}(\omega,z)\otimes
	\Bigg[ 
		[V^{(0)}(z,x)]_+ \ln\frac{Q^2}{\mu^2}-
	\frac{3}{2}[V^{\rm b}(z,x)]_+ -\frac{3}{2}C_F\delta(z-x)+
\nonumber\\
&&{\cal P}[g(z,x)]_+ \Bigg] +
{\cal P}\hat{T}^{(0)}(\omega,z)\ln\left\{1+\omega(\bar{z}-z)\right\}
						\otimes [V^{(0)}(z,x)]_+ +
\\
&&\hspace{-1.3mm}(I-{\cal P})\left[
	\hat{T}^{(0)}(\omega,z)\ln\left\{1+\omega(\bar{z}-z)\right\} -
	\hat{T}^{(0)}(\omega,y)\otimes S(y,z)
\right]\otimes [V^{(0)}(z,x)]_+.
\nonumber
\end{eqnarray}
The first two lines contain only diagonal terms, which provide the
$\alpha_s$-corrections to the overall normalization of the Wilson
coefficients
for the conformally covariant OPE. The off-diagonal terms in the last
line generate the shift of the canonical dimension by the anomalous one.
This can be seen by a straightforward calculation:
\begin{eqnarray}
 (I-{\cal P})\left[
	\hat{T}^{(0)}\ln\left\{1+\omega(\bar{z}-z)\right\} -
	\hat{T}^{(0)}\otimes S\right]\otimes[V^{(0)}]_+\otimes
 {(1-x)x \over N_k} C_k^{3\over 2}(2x-1)=
\\
{\gamma_k^{(0)}\over 2} \frac{(2\omega)^k}{(1+\omega)^{k+1}} \frac{d}
{d\rho} (1+\omega)^{-\rho} {_{2}F}_1\left({k+1+\rho, k+2+\rho \atop
2(k+2+\rho)}\Bigg|\frac{2\omega}{1+\omega}\right)_{|\rho=0}.
\nonumber
\end{eqnarray}
Therefore, this term coincides with the conformal prediction 
for the structure of the Wilson coefficients (\ref{conf-T-def-coe-QCD}).

It remains to be shown that the
normalization is consistent with the NLO calculation of the non-singlet
sector for the polarized structure function $g_1$ measured in DIS.
It is known that the diagonal part of the pion evolution
kernel coincides with the non-singlet splitting kernel. This was
analytically shown in
\cite{GeyDitHorMueRob88,MueRobGeyDitHor94} by taking the limit of an
extended pion evolution kernel:
\begin{eqnarray}
\label{TFF-Lim}
P(z)=\lim_{\eta\to 0} \frac{1}{|2\eta|}\gamma\left(\frac{z}{\eta},
\frac{1}{\eta}\right),
\mbox{\ }
	\gamma\left(t,t'\right)=
	V\left(\frac{1+t}{2},\frac{1+t'}{2}\right)_{|
\theta(t-t') \to \epsilon(1-t) \theta\left(\frac{1-t}{1-t'}\right)
\theta\left(\frac{t-t'}{1-t'}\right)}.
\end{eqnarray}
The extension of $V\left(t,t'\right)$ into the
whole $t,t'$-plane is unique and it is  done in practice by replacing the
corresponding $\theta$-functions.

As mentioned before, the off-diagonal part in Eq.\
(\ref{TFF-HarSca-NLO-mod}) does not contribute to the forward case. For the
terms of the diagonal part that are given as a convolution with $T^{(0)}$
the procedure (\ref{TFF-Lim}) provides the following NLO corrections:
\begin{eqnarray}
\label{TFF-Coe-Con-1}
C_F\int_0^1 dx\left(\left[\frac{1+x^2}{1-x}\right]_+
	\ln\frac{Q^2}{\mu^2}-
	\frac{3}{2}\left[\frac{2x}{1-x}\right]_+ -\frac{3}{2}\delta(1-x)-
   2\left[\frac{\ln(1-x)}{1-x}\right]_+ \right) x^k.
\end{eqnarray}
Convoluting the remaining term,
${\cal P}\hat{T}^{(0)}(\omega,z)\ln\left\{1+\omega(\bar{z}-z)\right\}\otimes
[V^{(0)}(z,x)]_+$,
first with the Gegenbauer polynomials and then extracting the diagonal
part gives:
\begin{eqnarray}
\label{TFF-Coe-Con-2}
-C_F \left[\psi(k+1)-\psi(1)\right]
\left[\frac{3}{2}+\frac{1}{(k+1)(k+2)}-2\psi(k+2)+2\psi(1)\right].
\end{eqnarray}
Putting (\ref{TFF-Coe-Con-1}) and (\ref{TFF-Coe-Con-2}) together,
the whole NLO contribution to the overall normalization follows:
\begin{eqnarray}
\label{TFF-Coe-Con-all}
e_k &=&
1+\frac{\alpha_s}{2\pi}C_F
\Bigg(
	\ln\frac{Q^2}{\mu^2}
	\left[
		\frac{3}{2}+\frac{1}{(k+1)(k+2)}-2S_1(k+1)
	\right] + 	2S_{1,1}(k)-
\\
&&\hspace{2cm}2S_{2}(k)+\left[
		\frac{1}{k+1} + \frac{1}{k+2} +\frac{3}{2}
	\right]S_1(k) +
	\frac{3}{k+1} - \frac{9}{2}
\Bigg),
\nonumber
\end{eqnarray}
where $S_m(k)=\sum_{i=1}^k (1/i)^m$
and  $S_{m,n}(k)=\sum_{i=1}^k (1/i)^m S_n(i)$.
Taking into account the different definition of moments in DIS, i.e.\
$k\to k-1$, the obtained normalization (\ref{TFF-Coe-Con-all}) coincides
with the Wilson coefficients in longitudinal polarized DIS computed in
Ref.\ \cite{ZijNee94} in the 't Hooft--Veltman--Breitenlohner--Maison (HVBM)
scheme\footnote{In the HVBM scheme the anticommutativity
of $\gamma^5$ in the non-singlet sector is restored effectively
by a finite renormalization. In this case, it is equivalent to the naive
$\gamma^5$ prescription, which was used for the calculation of the
NLO correction to the pion transition form factor.}
\cite{HooVel72,BreMai77}.

Finally, we show that also the NLO calculation for the transverse helicity
amplitude $T_{++}$ in the c.m.\ frame of the transition form factor for the
non-singlet scalar mesons \cite{AguCha81} coincides with the NLO corrections
to unpolarized DIS. It is sufficient to consider the difference to the
pseudo-scalar case, which can be written as a convolution of the hard
scattering part with a diagonal kernel \cite{AguCha81}:
\begin{eqnarray}
\hat{T}^{(0)}(\omega,z)\otimes
	\frac{\alpha_s}{2\pi}\left( [V^{(0)}(z,x)]_+ -[V^{\rm b}(z,x)]_+
- \frac{C_F}{2} \delta(z-x) \right),
\end{eqnarray}
which has the eigenvalues $\frac{\alpha_s}{2\pi}C_F/((k+1)(k+2))$. Thus, the
only difference to the pseudo-scalar case appears in the normalization given
by these eigenvalues. In the notation of the DIS hadronic tensor, the
considered helicity amplitude corresponds to the generalization of $F_1$ to
non-forward processes. The difference of the corresponding DIS Wilson
coefficient \cite{NeeZij91,ZijNee91a,ZijNee92} and $e_k$ in Eq.\
(\ref{TFF-Coe-Con-all}) is precisely $\frac{\alpha_s}{2\pi}C_F/((k+1)(k+2))$
in NLO.

\subsection{A first view beyond NLO}

It has been shown to LO that the measured pion transition form factor at
large momentum transfer \cite{CELLO91,Sav97}, where one photon is almost
real, supports the asymptotic distribution amplitude or even more narrow
ones \cite{Rad94,JakKroRau96,Ong95}. So it is phenomenologically very
interesting to study the higher-order corrections to this distribution
amplitude. In LO the asymptotic distribution amplitude $\varphi^{\rm
as}(x)=6x(1-x)$ does not evolve with $Q^2$, but it is well-known that this
property is spoiled in the $\overline{\mbox{MS}}$ scheme to NLO
\cite{MikRad86,Mue95} by the mixing of the operators. In the conformal limit
of the theory the conformal normalization conditions restore the
non-evolution of the asymptotic distribution amplitude. The pion transition
form factor for this amplitude is given by the first term of the conformal
OPE (\ref{conf-T-def}):
\begin{eqnarray}
\label{TFF-Pre-conOPE}
Q^2 F(\omega,Q^2) &=& {\sqrt{2} f_\pi\over 3} 
	{2 \over 1+\omega}\,
	{_{2}F}_1\left({1,2 \atop 4}\Bigg|\frac{2\omega}{1+\omega}\right)
	c_0(\alpha_s),
\end{eqnarray}
where the expectation value of the first operator, given by the axial current,
provided the pion decay constant $f_\pi=130.7 \mbox{ MeV}$. The coefficient
$c_0(\alpha_s)$ is normalized to 1 at LO. For the case that one
photon is almost real, i.e. $\omega=1$, we get 
\begin{eqnarray}
\label{TFF-Pre-conOPE-asy}
Q^2 F(1,Q^2) = \sqrt{2} f_\pi c_0(\alpha_s) =
0.185\,  c_0(\alpha_s) \mbox{\ GeV}.
\end{eqnarray}
The predictive power of the conformal OPE tells us that the  coefficient
$c_0(\alpha_s)$ is the value
 of the Bjorken sum rule, which is calculated up to order $\alpha_s^3$
\cite{GorLar86,LarVer91}. For three active flavours the
numerical result reads\footnote{The $\alpha_s^4$-correction has been
estimated to be negative too \cite{KatSta95,SamEllKar95}.} 
\begin{eqnarray}
\label{BjoSumRul}
c_0(\alpha_s)=
	1-\frac{\alpha_s}{\pi}-3.58333\left(\frac{\alpha_s}{\pi}\right)^2-
	20.21527 \left(\frac{\alpha_s}{\pi}\right)^3+O\left(\alpha_s^4\right).
\end{eqnarray}
Now we can give a rough estimate of the higher-loop corrections, which
reduce the LO prediction at a scale of $Q^2=2\mbox{\ GeV}^2$, where
$\alpha_s$ is assumed to be $0.35$, by about 18\%, coinciding very well with
the experimental results at this scale \cite{CELLO91,Sav97}. Note that the
$\alpha_s^2$ correction to the coefficient function of $g_1$ is given in
Ref.\ \cite{ZijNee94} and, therefore, the next-to-next-to-leading order (NNLO)
prediction for the photon-to-pion transition form factor can be also given
in the conformal limit for arbitrary DA's.

Now let us consider the effects coming from the conformal symmetry breaking,
which is manifested in the off-diagonal part of the Wilson coefficients and
the anomalous dimension matrix. In the conformally covariant subtraction
scheme considered here these terms are induced by the renormalization of the
coupling and have to be proportional to the $\beta$-function. In NLO only
the first coefficient of $\beta/g=-\alpha_s/(4\pi)\beta_0+O(\alpha_s^2)$
with $\beta_0=11-2n_f/3 $ enters in the anomalous dimension matrix (or in
the evolution kernel) and the off-diagonal term related to it was correctly
predicted by the conformal constraints \ref{conf-constr-KD}. On the other
hand it is obvious that in this order, this off-diagonal term can
be simply calculated from the two-loop diagram contributing the
$n_f$-dependent part to the gluon vacuum polarization (quark bubble)
\cite{Mik97,GodKiv97}. In NNLO terms proportional to the $\beta$-function
appear in both the Wilson coefficients and the evolution kernel. While the
off-diagonal part to the Wilson coefficients can be obtained in the same
manner as described, the special conformal anomaly should be known in order
$\alpha_s^2$ to treat the evolution of the DA in the correctly.

Although, the $n_f$-dependent part of the Wilson coefficients in NNLO can be
obtained from the result given in \cite{GodKiv97,BelSch98}, we only discuss
here the conformal symmetry breaking in NLO. Because of this breaking the
asymptotic distribution amplitude will evolve also in the conformal
subtraction scheme. Thus, let us first study the fixed $\alpha_s$ regime,
where a term proportional to $\beta_0$ arises in the $\alpha_s$ corrections
to the eigenfunctions (\ref{efcoef}) responsible for the evolution of the
asymptotic distribution amplitude. A renormalization group transformation
absorbs this term into the hard scattering part:
\begin{eqnarray}
\label{c0-fix-alp}
c^{\rm fix}_0(\alpha_s)=
	1-\frac{\alpha_s}{2\pi}\left(2+\frac{\beta_0}{6} \right) +
 O\left(\alpha_s^2\right).
\end{eqnarray}
Note that this $\beta_0$ term was predicted from the $\beta$-function by
the conformal constraint (\ref{conf-constr-KD}) and represents the first term
of the series $\beta/g\left(1/3+O(\alpha_s)\right)$. For the above values
the $\beta_0$ term provides a reduction of about 8\%, so that the whole net
reduction in NLO is of about 19.5\%. Because of a partial cancellation
between the different conformal symmetry breaking terms \cite{Mue95} for the
lowest moments, this reduction is similar to that in the
$\overline{\mbox{MS}}$ scheme, where the correction to the hard scattering
amplitude is $1-5 \alpha_s/(3\pi)$.

Now we consider the real case, where the coupling constant is running. Then,
the evolution of the asymptotic distribution amplitude is only avoidable if
the matrix $\hat{B}(\alpha_s)$, which diagonalizes the anomalous dimension
matrix is renormalization-group-invariant, i.e. $\hat{B}(\alpha_s(\mu),\mu)$
depends explicitly on $\mu$. Here we proceed in the manner proposed in
\cite{BroDamFriLep86,MikRad86}, which was explored in more detail in
\cite{Mue95}. Note that already a renormalization group transformation was
done to diagonalize the kernel for fixed $\alpha_s$, so that the evolution
kernel is different from that in the $\overline{\mbox{MS}}$ scheme.
Generally, the non-perturbative input can be taken from sum rules, lattice
or model calculations at a lower scale. Assuming that the ``input" $6x(1-x)$
is given at a scale of $Q \sim 1 \mbox{\ GeV}$ the evolution provides an
additional negative effect of almost 2\% for $Q^2=2\mbox{\ GeV}^2$ and of
almost 3.5\% for $Q^2=8\mbox{\ GeV}^2$. The resulting prediction is $0.148$
at $Q^2=2\mbox{\ GeV}^2$. A more detailed analysis including other
distribution amplitudes will be given elsewhere.

\section{Conclusion}

In this article we reviewed an appropriate technique, developed previously,
based on the true conformal Ward identities and conformal constraints, to
analyse conformal symmetry breaking in a massless quantum field theory due
to the renormalization of the UV-divergences. Since we are dealing with Ward
identities for the basis fields to define the anomalous terms of
gauge-invariant operators the conformal symmetry is also spoiled by the
gauge fixing. However, finally we are interested in predicting physical
quantities from which these terms should be absent. This point of view allows us
to understand conformal symmetry and its breaking in quantum field theories
without further conformal assumptions that led in the past to conflicts
between conformal predictions and explicit calculations. This approach is
also sufficient to reanalyse more directly the failure of the conformal
prediction from the light-cone expansion of a non-local operator for the
eigenfunctions of the pion evolution kernel in gauge field theory.

We employed this technique to prove that a factorization scheme exists in
which the conformal covariance of composite operators holds true in the
conformal limit of the theory. The transformation from an arbitrary scheme
to the conformal scheme is given by the special-conformal anomaly matrix.
Consequently, the essential assumption to construct the conformally
covariant OPE of two local currents can be fulfilled by requiring
appropriate normalization conditions. This OPE provides powerful
scheme-independent predictions, which were used for exclusive two-photon
processes in the generalized Bjorken region; also, restricted to the
conformal limit, for simplicity we did not consider the full kinematical
structure of the hadronic tensor. Since these predictions are
scheme-independent they hold true in any scheme; however, it is not a
trivial task to see this in the explicit calculated expressions beyond the
LO.

At this stage it is not clear how to obtain, in an economical way, the terms
proportional to the $\beta$-function that are missing in the conformal
limit. It seems to be worthwhile studying if the non-decomposable
irreducible representations that allow us to include the $\beta$-function in
the conformal symmetry interpretation also have predictive power. This would
avoid having to formally rely on the hypothetical conformal limit. A second
point of view is to use the common irreducible representations and to
consider the conformal symmetry-breaking terms as perturbation proportional
to the $\beta$-function. It is very interesting that in the case of the
Crewther relation the $\beta$-function can be absorbed by the BLM
scale-fixing prescription into the scale of the coupling constant
\cite{BroLu95,BroGabKatLu96,Rat96}.

In NLO conformal symmetry-breaking terms do not appear in the Wilson
coefficients. The $\beta$-dependence of the anomalous-dimension matrix is
predicted by the conformal constraints for the dimensional regularized
theory in the $\overline{\rm MS}$ scheme; however, the renormalization group
transformation to the conformal scheme provides an additional
$\beta$-dependence of the anomalous-dimension matrix in terms of the
special-conformal anomaly matrix computed in the $\overline{\rm MS}$ scheme.
This can be applied in a straightforward manner to predict the evolution in
the singlet channel to NLO only by a one-loop calculation of the
special-conformal anomaly matrix.

\section*{Acknowledgements}

I would like to thank S. Brodsky, R. Crewther and A.L. Kataev for
discussions, which inspired me to study the conformal properties of the OPE
in QCD. The author is grateful to the CERN Theory Division for its
hospitality during his visit, where this work was done. He was financially
supported by the Deutsche Forschungsgemeinschaft (DFG).

\appendix
\section*{}
\label{App-B-pro}

Here we carry out the calculation providing  Eq.\ (\ref{conAno-for-coO}).
For this reason, we first show the identity:
\begin{eqnarray}
\label{B-prop}
{}\left[\hat{a},\hat{B}\right]_{kk'}= a(k,k') B_{kk'} =
 -\left\{\hat{\gamma}^c(k')\hat{B}\right\}_{kk'}, \quad
a(k,k')=2(k-k')(k+k'+\nu_1+\nu_2).
\end{eqnarray}
Let us mention that by iteration and taking into account $B_{kk}=1$,
Eq.\ (\ref{blefdt-1}) follows. In accordance with the definition
(\ref{sol-transformation-1}) of the matrix $\hat{B}$ we introduce the
notation
\begin{eqnarray}
\label{B-prop-def}
B_{kk'}=\sum_{i=0}^\infty \left\{\hat{\Gamma}(k')^{i}\right\}_{kk'}, \quad
\hat{\Gamma}^{0}=\hat{1}, \mbox{\ and\ }
\Gamma_{mn}(k')=
{\cal L}_{k'} \gamma^{\rm ND}_{mn}=
- \frac{\gamma^{\rm ND}_{mn}}{\gamma_{m}-\gamma_{k'}}
\end{eqnarray}
as well as the inverse operator ${\cal L}_{k'}^{-1}$:
${\cal L}_{k'}^{-1} \gamma^{\rm ND}_{mn}=
-\left(\gamma_{m}-\gamma_{k'}\right) \gamma^{\rm ND}_{mn}$. From 
the conformal constraints (\ref{conf-constr-KD-1}) we obtain formally:
\begin{eqnarray}
\label{B-prop-der-1}
\left[\hat{a},\hat{\Gamma}(k')\right]=
-\hat{\gamma}^{c}(l)+{\cal L}_{k'} \hat{K}(l) +
 {\cal L}_{k'}\hat{\gamma}^{c}(l){\cal L}_{k'}^{-1},
\end{eqnarray}
where $ \hat{K}(l)=\left[\hat{\gamma}^{\rm ND},\hat{\gamma}^c(l)\right]$.
Note that the $l$-independence of the r.h.s.\ is ensured by the constraints
(\ref{conf-constr-KP}) and that the last term on this side induces for
$\left[\hat{a},\hat{\Gamma}(k')\right]_{kk'}$ the  contribution
$\left\{{\cal L}_{k'}
	\hat{\gamma}^{c}(l){\cal L}_{k'}^{-1}\hat{1}\right\}_{kk'}
=\hat{\gamma}^{c}_{kk'}(l) \delta_{kk'}$, so that  
$\left[\hat{a},\hat{\Gamma}(k')\right]_{k'k'}=0$ is identically
fulfilled for $k=k'$. Repeated application of
\begin{eqnarray}
\left[\hat{a},\hat{\Gamma}(k')^{i-1}\hat{\Gamma}(k')\right]=
\left[\hat{a},\hat{\Gamma}(k')^{i-1}\right]\hat{\Gamma}(k') +
\hat{\Gamma}(k')^{i-1}\left[\hat{a},\hat{\Gamma}(k')\right]
\nonumber
\end{eqnarray}
then implies the following form:
\begin{eqnarray}
\label{B-prop-der-2}
\left[\hat{a},\hat{\Gamma}(k')^i\right]=
-\hat{\gamma}^{c}(l)\hat{\Gamma}(k')^{i-1} -
 \hat{R}^{(i-1)}(k',l)+ \hat{R}^{(i)}(k',l) +
 \hat{\Gamma}(k')^{i-1}
{\cal L}_{k'}\hat{\gamma}^{c}(l){\cal L}_{k'}^{-1},
\end{eqnarray}
with $\hat{R}^{(-1)}=0$. For
$\left[\hat{a},\hat{\Gamma}(k')^i\right]_{kk'}$ the last term in the
r.h.s.\ is proportional to $\gamma^c(l)_{k'k'}$ and, because of
the property $\gamma^c(k')_{k'k'}=0$,  it can be avoided
for $l=k'$. Obviously, employing Eq.\ (\ref{B-prop-der-2})
we get the identity (\ref{B-prop}):
\begin{eqnarray}
\label{B-prop-der-3}
\left[\hat{a},\hat{B}\right]_{kk'}&=&
\sum_{i=0}^\infty \left[\hat{a},\Gamma(k')^{i}\right]_{kk'}
=
-\sum_{i=1}^\infty\left(
\left\{\hat{\gamma}^{c}(k')\hat{\Gamma}(k')^{i-1}\right\}_{kk'} +
 \hat{R}^{(i-1)}_{kk'}- \hat{R}_{kk'}^{(i)} \right)
\nonumber\\
&=&-\sum_{i=1}^\infty
\left\{\hat{\gamma}^{c}(k')\hat{\Gamma}(k')^{i-1}\right\}_{kk'}
=-\left\{\hat{\gamma}^{c}(k')\hat{B}\right\}_{kk'}.
\end{eqnarray}

With the help of relation (\ref{B-prop}) the desired calculation is easy:
\begin{eqnarray}
\label{B-prop-der-4}
\left\{
	\hat{B}^{-1}
\left[\hat{a}(l)+\hat{\gamma}^c(l)\right]\hat{B}
\right\}_{kk'} &=&
\left\{	\hat{B}^{-1} \hat{B}\right\}_{kk'} a(k',l)+
\left\{\hat{B}^{-1}\left[\hat{a},\hat{B}\right] +
\hat{B}^{-1}\hat{\gamma}^c(l)\hat{B}\right\}_{kk'}
\nonumber\\
&=& a(k,l)\delta_{kk'} +
	\left\{\hat{B}^{-1}\left[\hat{\gamma}^c(l)-
	\hat{\gamma}^c(k')\right]\hat{B}\right\}_{kk'}
\\
&=& a(k,l)\delta_{kk'} +
	2(k'-l) \left\{\hat{B}^{-1}\hat{\gamma}\hat{B}\right\}_{kk'},
\nonumber
\end{eqnarray}
where we used $\hat{\gamma}^c(l)-\hat{\gamma}^c(k')=-2(l-k')\hat{\gamma}$.
Since the matrix $\hat{B}$ diagonalizes the anomalous-dimension matrix
the last line is identical with Eq.\ (\ref{conAno-for-coO}).


\begin{thebibliography}{10}

\bibitem{Cre72}
R. Crewther, Phys. Rev. Lett. {\bf 28},  1421  (1972).

\bibitem{AdlCalGroJac72}
S. Adler, C. Callan, D. Gross, and R. Jackiw, Phys. Rev. {\bf D6},  2982
  (1972).

\bibitem{LarVer91}
S. Larin and J. Vermaseren, Phys. Lett. {\bf B259},  345  (1991).

\bibitem{GorKatLar91}
S. Gorishny, A. Kataev, and S. Larin, Phys. Lett. {\bf B259},  144  (1991).

\bibitem{Cre97}
R. Crewther, Phys. Lett. {\bf B397},  137  (1997).

\bibitem{BroKat93}
D. Broadhurst and A. Kataev, Phys. Lett. {\bf B315},  179  (1993).

\bibitem{BroLu95}
H. Lu and S. Brodsky, Phys. Rev. {\bf D48},  3310  (1995).

\bibitem{Rat96}
J. Rathsman, Phys. Rev. {\bf D54},  3420  (1996).

\bibitem{FerGriGat71}
S. Ferrara, , R. Gatto, and A. Grillo, Phys. Lett. {\bf 36B},  124  (1971),
  erratum {\bf 38B}, 188 (1972).

\bibitem{FerGriGat71a}
S. Ferrara, R. Gatto, and A. Grillo, Nucl. Phys. {\bf B34},  349  (1971).

\bibitem{FerGriGat72a}
S. Ferrara, R. Gatto, and A. Grillo, Phys. Rev. {\bf D5},  3102  (1972).

\bibitem{BonSarTon72}
L. Bonora, G. Sartori, and N. Tonin, Nuovo Cimento {\bf A10},  667  (1972).

\bibitem{CraDobTod85}
N. Craigie, V. Dobrev, and I. Todorov, Ann. Phys. {\bf 159},  411  (1985).

\bibitem{FerGriPar73}
S. Ferrara, A. Grillo, and G. Parisi, Nucl. Phys. {\bf B54},  552  (1973).

\bibitem{BroFriLepSac80}
S. Brodsky, Y. Frishman, G. Lepage, and C. Sachradja, Phys. Lett. {\bf 91B},
  239  (1980).

\bibitem{DitRad84}
F.-M. Dittes and A. Radyushkin, Phys. Lett. {\bf 134B},  359  (1984).

\bibitem{Sar84}
M. Sarmadi, Phys. Lett. {\bf 143B},  471  (1984).

\bibitem{Kat85}
G. Katz, Phys. Rev. {\bf D31},  652  (1985).

\bibitem{MikRad85}
S. Mikhailov and A. Radyushkin, Nucl. Phys. {\bf B254},  89  (1985).

\bibitem{BroDamFriLep86}
S. Brodsky, P. Damgaard, Y. Frishman, and G. Lepage, Phys. Rev. {\bf D33},
  1881  (1986).

\bibitem{Mue91a}
D. M{\"u}ller, Z. Phys. {\bf C49},  293   (1991).

\bibitem{Mue94}
D. M{\"u}ller, Phys. Rev. {\bf D49},  2525   (1994).

\bibitem{MueRobGeyDitHor94}
D. M{\"u}ller {\it et~al.}, Fortschr. Phys. {\bf 42},  101  (1994).

\bibitem{Ji96}
X. Ji, Phys. Rev. Lett. {\bf 78},  610  (1997).

\bibitem{Ji97}
X. Ji, Phys. Rev. {\bf D55},  7114  (1997).

\bibitem{Rad96}
A. Radyushkin, Phys. Lett. {\bf B380},  417  (1996).

\bibitem{Pluto84}
C.~Berger et~al. (PLUTO~Collaboration), Phys. Lett. {\bf B142},  125  (1984).

\bibitem{TPC90}
H.~Aihara et~al. (TPC/Two Gamma~Collaboration), Phys. Rev. Lett. {\bf 64},  172
  (1990).

\bibitem{CELLO91}
H.-J.~Behrend et~al. (CELLO~Collaboration), Z.\ Phys. {\bf C49},  401  (1991).

\bibitem{Sav97}
J.~Gronberg et~al. (CLEO~Collaboration), Phys. Rev. {\bf D57},  33  (1998).

\bibitem{MacSal69}
G. Mack and A. Salam, Ann. Phys. {\bf 53},  174  (1969).

\bibitem{Mak81}
Y. Makeenko, Sov. J. Nucl. Phys. {\bf 33},  440  (1981).

\bibitem{Ohr82}
T. Ohrndorf, Nucl. Phys. {\bf B198},  26  (1982).

\bibitem{BalBra89}
I. Balitsky and V. Braun, Nucl. Phys. {\bf B311},  541  (1989).

\bibitem{Zai88}
R. Zaikov, Lett. in Math. Phys. {\bf 16},  1  (1988).

\bibitem{Par72}
G. Parisi, Phys. Lett. {\bf 39B},  643  (1972).

\bibitem{Nie73}
N. Nielsen, Nucl. Phys. {\bf B65},  413  (1973).

\bibitem{Sar74}
S. Sarkar, Nucl. Phys. {\bf B83},  108  (1974).

\bibitem{Nie77}
N. Nielsen, Nucl. Phys. {\bf B120},  212  (1977).

\bibitem{AdlColDun77}
S. Adler, J. Collins, and A. Duncan, Phys. Rev. {\bf D15},  1712  (1977).

\bibitem{ColDunJog77}
J. Collins, A. Duncan, and S. Joglekar, Phys. Rev. {\bf D16},  438  (1977).

\bibitem{Min76}
P. Minkowski, Berne preprint-76-0813 (unpublished).

\bibitem{Cha80}
M. Chase, Nucl. Phys. {\bf B167},  125  (1980).

\bibitem{Cha80a}
M. Chase, Nucl. Phys. {\bf B174},  109  (1980).

\bibitem{GeyMueRob92}
B. Geyer, D. M{\"u}ller, and D. Robaschik,  in {\em Proceedings of the 1992
  Zeuthen Workshop on Elementary Particle Theory: ``Deep Inelastic
  Scattering''}, edited by J. Bl{\"u}mlein and T. Riemann
  (Teupitz/Brandenburg, Germany, 1992), Vol.~Nucl. Phys. B (Proc. Suppl.) 29A,
  pp.\ 22.

\bibitem{ColFraStr96}
J. Collins, L. Frankfurt, and M. Strikman, Phys. Rev. {\bf D56},  2982  (1997).

\bibitem{Rad96a}
A. Radyushkin, Phys. Lett. {\bf B385},  333  (1996).

\bibitem{GeyDitHorMueRob88}
F.-M. Dittes {\it et~al.}, Phys. Lett. {\bf 209B},  325  (1988).

\bibitem{JaiRal93}
P. Jain and J. Ralston,  in {\em Proceedings of the Workshop on Future
  Directions in Particle and Nuclear Physics at Multi-GeV Hadron Beam
  Facilities}, (Brookhaven National Laboratory,
  Upton, NY, 1993).

\bibitem{JiMelSon97}
X. Ji, W. Melnitchouk, and X. Song, Phys. Rev. {\bf D56},  5511  (1997).

\bibitem{NeeZij91}
W.~V. Neerven and E. Zijlstra, Phys. Lett. {\bf B272},  127  (1991).

\bibitem{ZijNee91a}
E. Zijlstra and W. van Neerven, Phys. Lett. {\bf B273},  476  (1991).

\bibitem{ZijNee92}
E. Zijlstra and W. van Neerven, Nucl. Phys. {\bf B383},  525  (1992).

\bibitem{ZijNee94}
E. Zijlstra and W. van Neerven, Nucl. Phys. {\bf B417},  61  (1994), erratum
   {\bf B426}, 245 (1994).

\bibitem{BroLep80}
S. Brodsky and G. Lepage, Phys. Rev. {\bf D22},  2157  (1980).

\bibitem{AguCha81}
F.~D. Aguila and M. Chase, Nucl.\ Phys. {\bf B193},  517  (1981).

\bibitem{Bra83}
E. Braaten, Phys. Rev. {\bf D28},  524  (1983).

\bibitem{KadMikRad86}
E. Kadantseva, S. Mikhailov, and A. Radyushkin, Sov. J. Nucl. Phys. {\bf 44},
  326  (1986).

\bibitem{Mue95}
D. M{\"u}ller, Phys. Rev. {\bf D51},  3855   (1995).

\bibitem{HooVel72}
G. 't~Hooft and M. Veltman, Nucl. Phys. {\bf B44},  189  (1972).

\bibitem{BreMai77}
P. Breitenlohner and D. Maison, Commun. Math. Phys. {\bf 52},  11  (1977).

\bibitem{Rad94}
A. Radyushkin, CEBAF-TH-94-15 (unpublished).

\bibitem{JakKroRau96}
R. Jakob, P. Kroll, and M. Raulfs, J. Phys. {\bf G22},  45  (1996).

\bibitem{Ong95}
S. Ong, Phys. Rev. {\bf D52},  3111  (1995).

\bibitem{MikRad86}
S. Mikhailov and A. Radyushkin, Nucl. Phys. {\bf B273},  297  (1986).

\bibitem{GorLar86}
S. Gorishny and S. Larin, Phys. Lett. {\bf B172},  109  (1986).

\bibitem{KatSta95}
A. Kataev and V. Starstenko, Mod. Phys. Lett. {\bf A10},  235  (1995).

\bibitem{SamEllKar95}
M. Samuel, J. Ellis, and M. Karliner, Phys. Rev. Lett. {\bf 74},  4380  (1995).

\bibitem{Mik97}
S. Mikhailov, Phys. Lett. {\bf B416},  421  (1997).

\bibitem{GodKiv97}
P. Gosdzinsky and N. Kivel, hep-ph/9707367 (unpublished).

\bibitem{BelSch98}
A. Belitsky and A. Sch\"afer, hep-ph/9801252 (unpublished).

\bibitem{BroGabKatLu96}
S. Brodsky, G. Gabadadze, A. Kataev, and H. Lu, Phys. Lett. {\bf B372},  133
  (1996).

\end{thebibliography}

\end{document}